**Evidence for Distributed Fault Energetics and Their Impact on Deformation in a Chemically Complex Alloy**

Short title: **Distributed Fault Energetics in Chemically Complex Alloy**

Kaijun Yin[1†], Jun-Ping Du[2†], Peijun Yu[2,3†], Rui Feng[4†], Hanyu Hou[1,5], Haw-Wen Hsiao[1], Ke An[4], Peter K. Liaw[6], Shigenobu Ogata[2*], and Jian-Min Zuo[7,8,1*]

[1]Department of Materials Science and Engineering and Materials Research Laboratory, The Grainger College of Engineering, University of Illinois at Urbana-Champaign, Urbana, IL, 61801, USA

[2]Department of Mechanical Science and Bioengineering, The University of Osaka, Osaka, 560-8531, Japan

[3]State Key Laboratory of Materials for Advanced Nuclear Energy, Shanghai University, Shanghai, 200444, China

[4]Neutron Scattering Division, Oak Ridge National Laboratory, Oak Ridge, TN, 37831, USA

[5]Center for Nanoscale Materials, Argonne National Laboratory, Lemont, IL, 60439, USA

[6]Department of Materials Science and Engineering, University of Tennessee, Knoxville, Knoxville, TN, 37996, USA

[7]Monash Center for Electron Microscopy, Monash University, VIC, Australia

[8]Department of Materials Science and Engineering, Monash University, VIC, Australia

[†]These authors contributed equally to this paper

[*]Corresponding authors: jianzuo@illinois.edu, ogata@me.es.osaka-u.ac.jp

**Abstract**

Chemically complex alloys (CCAs) feature intrinsically heterogeneous local chemical environments and, consequently, fluctuations in local fault energetics. However, experimentally quantifying their relationship remains challenging, leaving the role of this distributed energy landscape in deformation mechanisms incompletely resolved. Here, we develop a distribution-based framework linking experimentally measured stacking-fault widths to deformation-relevant apparent fault energy, revealing a distributed local fault-energy landscape in CrCoNi. The framework reveals the stabilizing role of energy fluctuations and captures an upward shift in the apparent fault energetics, from negative values toward zero following heat treatment, which atomistic simulations associate with the emergence of $L1_2$-type chemical short-range order (CSRO). Using one-dimensional kinetic Monte Carlo (1D-kMC) simulations, supported by electron microscopy observations, we further show that history-dependent changes in the local fault-energy landscape bias the competition among stacking faulting, nano-twinning and HCP transformation in CrCoNi. Our results provide an experimentally anchored, distribution-based framework for understanding deformation in CCAs as it evolves within a distributed fault-energy landscape shaped by local chemical order.

**Teaser**: Local chemical order shapes the statistical fault-energy landscape and deformation pathways in CrCoNi.

## INTRODUCTION

Deformation in chemically complex alloys (CCAs), including medium- and high-entropy alloys (M/HEAs) with multiple principal elements[1-4], remains incompletely understood despite the tremendous interest since their initial discovery[5,6]. Recent study has shown that the local atomic structure in CCAs deviates from ideal solid solutions through local chemical clustering and CSRO, which is strongly dependent on local composition and thermal history[7-10]. A key link between the local atomic structure and fault-mediated deformation is the stacking-fault energy (SFE), reflecting the energetic cost of forming a fault and its stability within the local chemical environment. In conventional face-centered cubic (FCC) alloys, SFE is often treated as an effective material parameter, leading to a well-defined separation between partial dislocations and an equilibrium stacking-fault (SF) width, which correlates with the relative propensity for dislocation slip, deformation twinning, and HCP transformation[11]. In CCAs, however, the SFE itself can be sensitive to the local chemical environment, including CSRO[9,12-17]. Pronounced lattice distortions from atom-to-atom variations in chemical bonding[18,19] also generate spatially heterogeneous slip resistance (Peierls stress or lattice friction)[20,21]. In such systems, fundamental understanding of deformation in CCAs requires the consideration of variations in SFEs and lattice frictions, more specifically, the treatment of deformation-relevant fault energetics as a statistical distribution rather than as a single uniform parameter. This distributed energy landscape is expected to fundamentally alter how dislocations nucleate, separate, and evolve, thereby shaping competing deformation pathways.

CrCoNi is representative of the challenges in studying fault energetics in CCAs[7-9,22-24]. Under mechanical loading, CrCoNi presents low-SFE-like deformation features with a propensity for nano-twinning and HCP transformation[25-28]. Theoretical studies have generally predicted negative or near-zero local SFEs in CrCoNi, with values that can shift with local chemical order. The calculated SFEs represent the intrinsic thermodynamic energy required to form a stacking fault in an atomic model[20,29,30]. In contrast, experimental measurements using the weak-beam dark-field (WBDF) imaging method[31,32] typically report positive apparent SFEs[9,25,33,34]. However, the reported values are an effective quantity, inferred indirectly from the equilibrium separation of partial dislocations through a force-balance model developed previously for conventional alloys. Thus, the calculated intrinsic SFEs and the WBDF measured and model interpreted quantities should not be considered as identical and compared directly[35].

Here, using the CrCoNi alloys processed by routes previously shown to produce distinct dominant CSRO motifs as model systems and a four-dimensional scanning transmission electron microscopy (4D-STEM) based experimental approach, we first show that the SF width is broadly distributed from few to ~$10^2$ nm. We then relate the measured SF-width distribution to a representative apparent SFE and distribution width, by introducing a statistical force-balance model for the analysis. To relate fault energetics to deformation pathways, we examine history-dependent generalized SFEs (GSFEs) and use 1D-kMC simulations to investigate the impact of CSRO. Finally, we provide in-situ neutron diffraction and ex-situ electron microscopy evidence for processing-dependent differences in faulting, nano-twinning, and HCP transformation.

## RESULTS

### *Experimental measurement of stacking fault widths and its distribution*

We started by preparing two CrCoNi alloys from the same cast, homogenized at 1,200 °C for 48 h, and then subjected to different final thermal processing conditions: one condition was water-quenched after homogenization (WQ), whereas the other was additionally aged at 1,000 °C for 120 h and furnace-cooled to room temperature (heat-treated, HT). These two processing routes were previously shown to produce predominant $L1_1$-type and $L1_2$-type CSRO motifs in WQ and HT CrCoNi, respectively[8]. We use these two states as experimentally motivated representations for different local chemical orders.

We measured SF widths and width distributions in the deformed WQ and HT samples using cepstral scanning transmission electron microscopy (STEM)[36]. Unlike WBDF imaging[31,32], which identifies paired partial dislocations through diffraction contrast and is most reliable in relatively uniform, lightly deformed regions, cepstral STEM is a four-dimensional STEM (4D-STEM) method that captures a diffraction pattern at every probe position and uses cepstral analysis of diffraction patterns to map interatomic-vector signals (Fig. 1). Because it is diffraction-based, cepstral STEM enables sensitive mapping of SFs, nanotwins, and HCP lamellae even in heavily deformed regions (Suppl. Note 1). The cepstral STEM analysis was performed along the [110] zone axis to image SFs directly in an edge-on geometry. This approach enables measurements of the thickness-averaged SF width for both short and extended faults of mixed character (Fig. 1E; Suppl. Notes 2). Cepstral STEM also images the faulted region itself rather than only the bounding partial dislocations, allowing SFs to be followed continuously in complex faulted substructures (Suppl. Note 1).

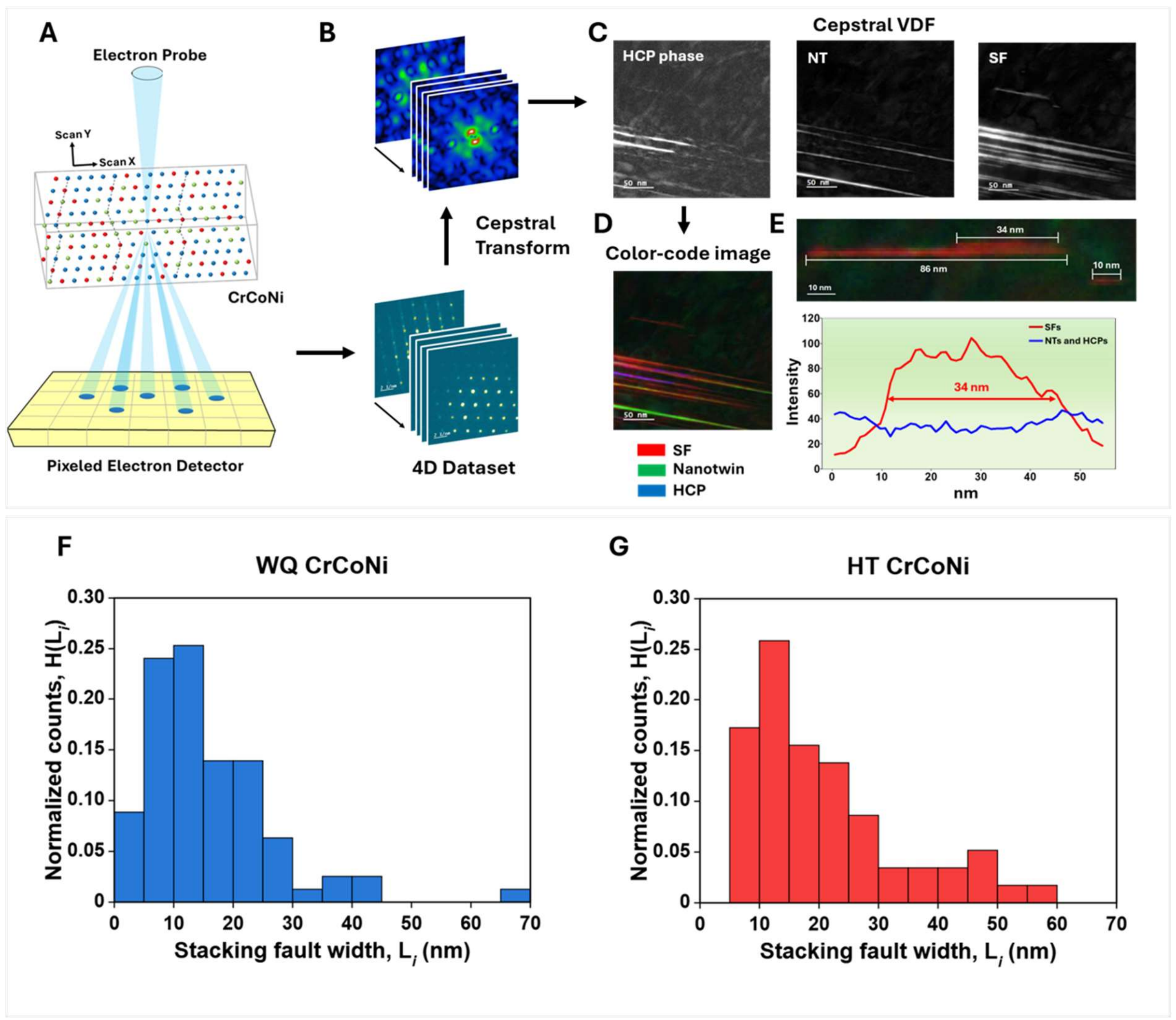

**Fig. 1 Cepstral STEM imaging of stacking faults and deformation substructures.** A) A 4D-STEM setup records an electron nanodiffraction pattern at every probe position. B) The diffraction dataset can be used for conventional STEM virtual dark-field (VDF) imaging from selected diffraction intensities or for cepstral VDF imaging using interatomic distance-sensitive signals. C-E) Compared with conventional STEM VDF imaging, cepstral VDF enables separate visualization of SFs, nanotwins (NTs), and HCP lamellae, with color-coded maps showing their spatial distributions. F,G) Experimental SF-width distributions in WQ and HT CrCoNi, plotted with $L$ ranging from 0 to 70 nm, respectively. The largest SF width that can be measured is limited by the field-of-view of 4D-STEM measurements, at ~260 nm.

Figures 1F,G show the measured SF-width distributions for the deformed WQ and HT samples, respectively, using cepstral STEM intensity profile (Fig. 1E). The measured SF width, $L$, is plotted ranging from a few nanometers up to 70 nm. These distributions are significantly broader than those previously reported (~20 nm or less) for the deformed CrCoNi alloys using WBDF imaging[9,13,20,25,33].

***Statistical inference of representative apparent fault energetics and distribution width in the water-quenched and heat-treated samples***

To account for the broad distribution of SF widths, we developed a statistical model that incorporates an "apparent" local SFE that combines the lattice friction force ($f_0$) with the local SFE ($E_{\mathrm{LSF}}$), $E_{\mathrm{LSF}}^{\mathrm{app}} = E_{\mathrm{LSF}} + f_0$, and considers its local fluctuations (Fig. 2A). $E_{\mathrm{LSF}}^{\mathrm{app}}$ represents the net force per unit length on a partial dislocation. As in pure metals, mechanical equilibrium requires that the net force on each partial dislocation vanishes at the equilibrium SF width $L$[20,37], such that:

$$L = \frac{f(G, b_{\mathrm{p}}\nu, \beta)}{E_{\mathrm{LSF}}^{\mathrm{app}}} \tag{1}$$

where $f(G, b_{\mathrm{p}}\nu, \beta) = \frac{Gb_{\mathrm{p}}^2}{8\pi}\left(\frac{2-\nu}{1-\nu}\right)\left(1 - \frac{2\nu\cos(2\beta)}{2-\nu}\right)$ depends on the shear modulus $G$, the Burgers vector magnitude $b_{\mathrm{p}}$ of the partials, Poisson's ratio $\nu$ , and the angle $\beta$ between the full Burgers vector and the dislocation line. The probability density function (PDF) of $E_{\mathrm{LSF}}^{\mathrm{app}}$ is assumed to be Gaussian, ~$N(E_{\mathrm{SF}},\sigma^2)$, with the mean $E_{\mathrm{SF}}$ and variance $\sigma^2$ (Suppl. Note 3), which gives

$$g\left(E_{\mathrm{LSF}}^{\mathrm{app}}\right) = \frac{1}{\sqrt{2\pi}\sigma} e^{-\frac{\left(E_{\mathrm{LSF}}^{\mathrm{app}} - E_{\mathrm{SF}}\right)^2}{2\sigma^2}}. \tag{2}$$

Here, $g(E_{\mathrm{LSF}}^{\mathrm{app}})$ represents the probability that a partial dislocation experiences a particular value of $E_{\mathrm{LSF}}^{\mathrm{app}}$ at its local location. If two partials are separated by a distance $L$, and both experience the same positive value of $E_{\mathrm{LSF}}^{\mathrm{app}} = \frac{f(G,b_{\mathrm{p}},\nu,\beta)}{L}$ required to satisfy the equilibrium force balance, and the probability density of this event is of $g^2(E_{\mathrm{LSF}}^{\mathrm{app}}) / \int_0^{\infty} g^2(E_{\mathrm{LSF}}^{\mathrm{app}}) \mathrm{d}E_{\mathrm{LSF}}^{\mathrm{app}}$, when $L$ is sufficiently large (e.g. $L$ exceeding a few nanometers, which is on the order of the reported CSRO nanocluster size[8]). The probability of finding a SF width between $L$ and $L + \mathrm{d}L$ is given by $p(L)dL$, where:

$$p(L) = \frac{1}{\tilde{C}} e^{-\frac{(1/L - \widetilde{E_{\mathrm{SF}}})^2}{\tilde{\sigma}^2}} \frac{1}{L^2}, \tag{3}$$

and $\widetilde{E_{\mathrm{SF}}} = \frac{E_{\mathrm{SF}}}{f(G,b_{\mathrm{p}},\nu,\beta)}$, $\tilde{\sigma} = \frac{\sigma}{f(G,b_{\mathrm{p}},\nu,\beta)}$, $\tilde{C} = \frac{\tilde{\sigma}\sqrt{\pi}}{2}\left[1 + \mathrm{erf}\left(\frac{\widetilde{E_{\mathrm{SF}}}}{\tilde{\sigma}}\right)\right]$, $\mathrm{erf}(x)$ is the error function. Figure 2b shows $p(L)$ for the positive, zero, or negative values of $\widetilde{E_{\mathrm{SF}}}$ and a fixed $\tilde{\sigma}$. With increasing $L$, $p(L)$ rises from zero, reaches a peak, and then decays with a long tail, regardless of the sign of $\widetilde{E_{\mathrm{SF}}}$.

To determine the SFE ($E_{\mathrm{SF}}$) from the experimentally measured distribution of SF widths, we focus on the most probable SF width $L_{\mathrm{m}}$, and its corresponding maximum probability density, $P_{\mathrm{m}} = p(L_{\mathrm{m}})$. According to Equation (3), $L_{\mathrm{m}}$ is obtained by solving $\frac{dp(L)}{dL} = 0$, which yields:

$$L_{\mathrm{m}} = \frac{2}{\widetilde{E_{\mathrm{SF}}} + \sqrt{\widetilde{E_{\mathrm{SF}}}^2 + 4\tilde{\sigma}^2}}. \tag{4}$$

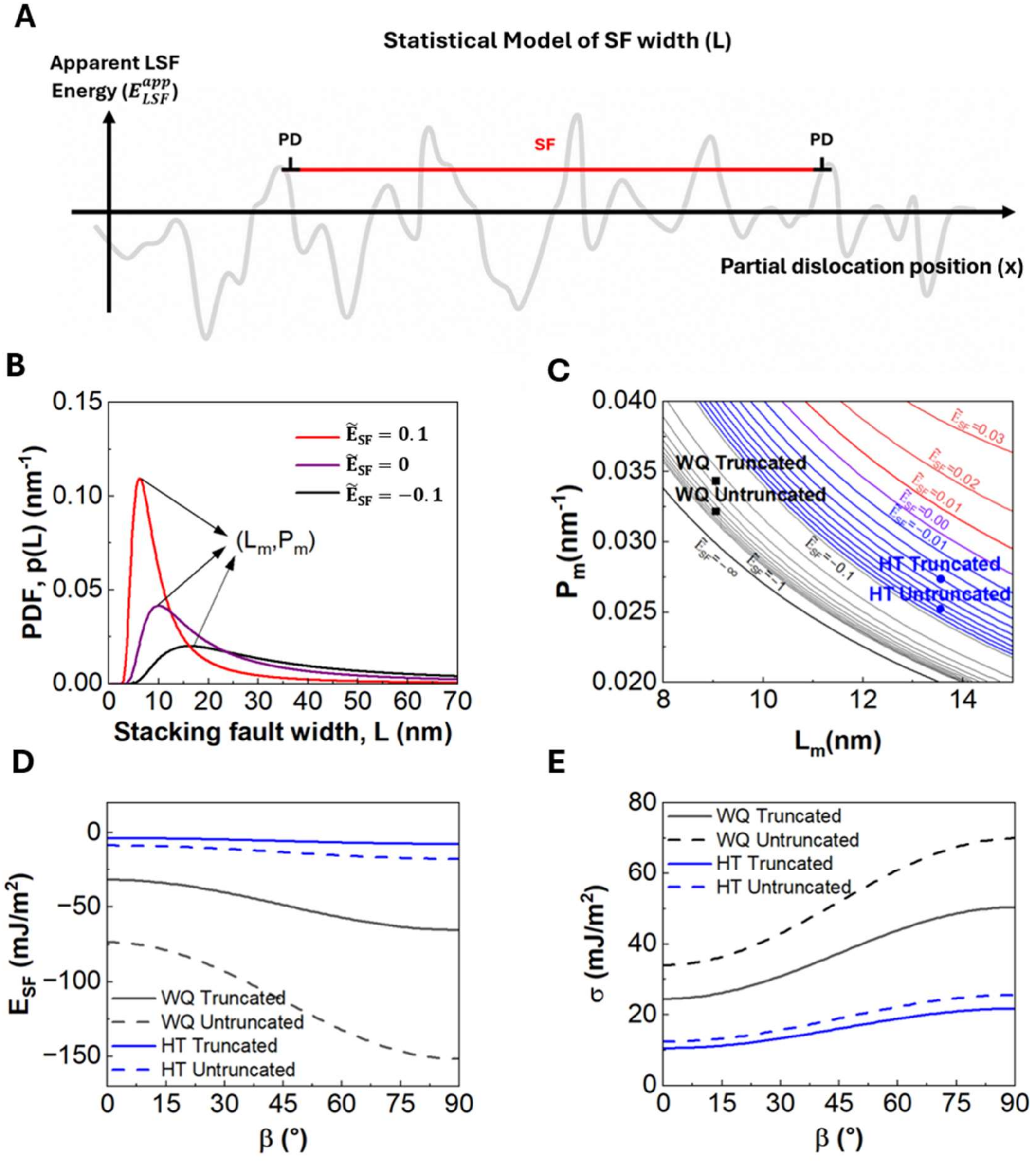

**Fig. 2 Theoretical probability model of SF-width distributions and comparison with experimental measurements.** A) Schematic illustration of the force balance between two partial dislocations (PDs) in a landscape of fluctuating apparent local SFEs and the resulting SF width $L$. B) Theoretical probability density functions (PDFs) of SF widths for different scaled apparent SFE values at fixed distribution width. Arrows mark the most probable SF width ($L_{\mathrm{m}}$) and its corresponding maximum probability density ($P_{\mathrm{m}}$). C) $L_{\mathrm{m}} - P_{\mathrm{m}}$ map for an FCC M/HEA. Solid lines represent constant scaled apparent SFE values with varying distribution width. Symbols show experimentally extracted ($L_{\mathrm{m}}, P_{\mathrm{m}}$) pairs from the measured SF-

width data using Gaussian kernel density estimation, with and without the truncation of $L$ at the observed maximum SF width (260 nm). D,E) Model-based apparent SFE and distribution width as functions of the angle $\beta$ between the full Burgers vector and the dislocation line, calculated using average values of $G$ and $\nu$ reported in Ref. 15.

Fig. 2C shows the relationship between $P_{\mathrm{m}}$ and $L_{\mathrm{m}}$ for a constant $\widetilde{E_{\mathrm{SF}}}$ and varying $\tilde{\sigma}$. In this plot, the boundary between positive and negative SFE values is defined by the curve at $E_{\mathrm{SF}} = \widetilde{E_{\mathrm{SF}}} = 0$, where the analytical form of the maximum probability density simplifies to

$$P_{\mathrm{m}} = \frac{\frac{2e^{-1}}{\sqrt{\pi}}}{L_{\mathrm{m}}} \approx \frac{0.415}{L_{\mathrm{m}}} \ . \tag{5}$$

In addition, a limiting boundary with $\widetilde{E_{\mathrm{SF}}} = -\infty$ exists at

$$P_{\mathrm{m}} = \frac{2e^{-2}}{L_{\mathrm{m}}} \approx \frac{0.271}{L_{\mathrm{m}}},$$

below which no $(L_{\mathrm{m}}, P_{\mathrm{m}})$ can appear.

The apparent SFE distribution was estimated from the experimentally measured SF-width distribution by extracting $(L_{\mathrm{m}}, P_{\mathrm{m}})$ using the Gaussian kernel density estimation (KDE) and bootstrap resampling (see Methods and Suppl. Note 4). The resulting $(L_{\mathrm{m}}, P_{\mathrm{m}})$ values for the WQ and HT samples are plotted in Fig. 2C. The conversion from $(L_{\mathrm{m}}, P_{\mathrm{m}})$ to apparent SFE depends on the dislocation character $\beta$ and elastic constants, the values reported here provide model-based apparent SFE ranges, direct measurements of a thermodynamic intrinsic SFE.

Results are shown both with and without truncation of the SF width at the observed maximum value (260 nm). In both analyses, the HT sample points are well-separated from the WQ sample points in Fig. 2C and shift toward contours corresponding to a higher apparent SFE, supporting the robustness of the relative upward shift in $\widetilde{E_{\mathrm{SF}}}$ against the uncertainties caused by the treatment of the large-width distribution tail.

Figures 2D and E show the apparent SFE and distribution width as functions of $\beta$. The absolute values were obtained from the normalized parameters using the average shear modulus $G$ and Poisson's ratio $\nu$ from Ref. [9]. The inferred apparent SFE range in the WQ state is negative, which shifts upward toward zero while remaining near zero or negative over the explored $\beta$ range, and the distribution width decreases in the HT state.

Thus, the statistical SF width distribution model emphasizes the relative shift between the WQ and HT states rather than a change in a single absolute SFE value, which is the most significant result despite the uncertainties related to the model and experimental measurements.

***Atomistic simulations of apparent local stacking fault energy distribution***

Previous atomistic simulations have demonstrated that CSRO transforms the intrinsic SFE from a static, composition-dependent value into a tunable landscape of $E_{\mathrm{LSF}}$ governed by local atomic environments[7,38,39]. However, apparent local fault energetics ($E_{\mathrm{LSF}}^{\mathrm{app}}$) relevant to deformation has not been considered. To gain the atomistic insight into the distribution of $E_{\mathrm{LSF}}^{\mathrm{app}}$ in CrCoNi, we performed simulations using a set of constructed atomic models: a WQ-like model, a random solid-solution model, and HT-like models with low, medium, and high degrees of $L1_2$-type ordering (Fig.3, Suppl. Note 5), following the protocols in Methods. The HT-like models vary in the spatial arrangement and degree of local ordering, including the cluster composition, volume fraction and pair-swap history. Thus, these models should be viewed as experimentally motivated representations of the WQ and HT states based on dominant $L1_1$- and $L1_2$-type motifs, rather than atomistic realizations of the measured specimens.

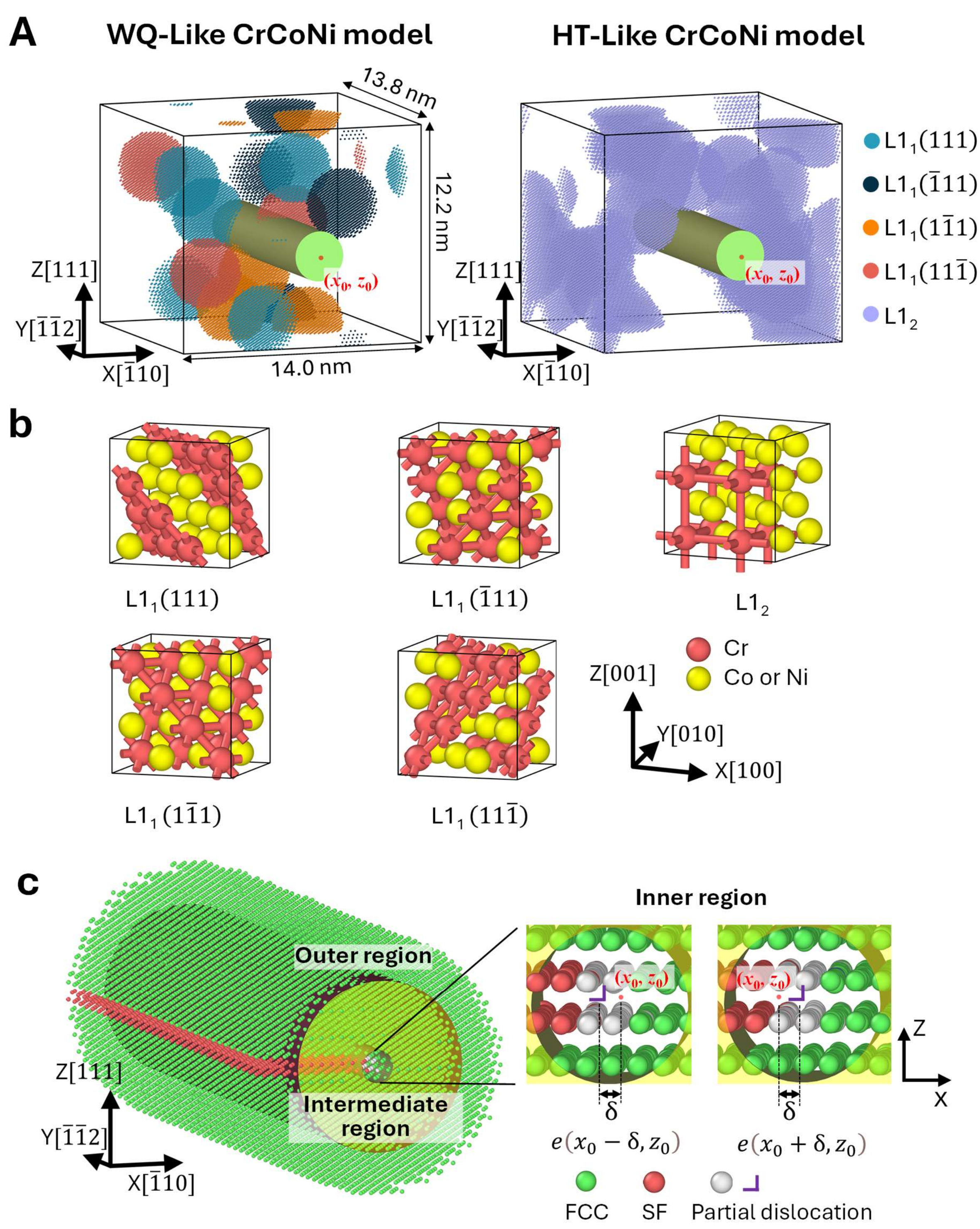

**Fig. 3 Atomistic modeling of the WQ and HT samples and the calculation of apparent local stacking-fault energy.** A) Schematic illustration of the WQ-like and HT-like CrCoNi atomistic models with the local CSRO highlighted in colors, along with the cylindrical volume used to calculate the apparent local stacking

fault energy $E_{\mathrm{LSF}}^{\mathrm{app}}(x_0, z_0)$. The cylinder is centered at ($x_0$, $z_0$) and aligned along the Y direction. The Miller indices following "$L1_1$" indicate the orientations of the Cr {111} close-packed planes. B) Atomistic structures of perfect $L1_1$ structure with the four different orientations of Cr {111} close-packed planes, and the $L1_2$ structure, which has a single Cr-Cr bond orientation along the <100> direction. Only nearest-neighbor Cr bonds are displayed. C) Atomistic representation of the cylindrical model, divided into inner, intermediate, and outer regions according to the constraints imposed during energy minimization. Here, $E_{\mathrm{LSF}}^{\mathrm{app}}(x_0, z_0) = \frac{e(x_0+\delta, z_0) - e(x_0-\delta, z_0)}{2\delta}$, where $e(x_0 + \delta, z_0)$ and $e(x_0 - \delta, z_0)$ denote the energies of the cylindrical model per unit dislocation line length for a partial dislocation positioned at $(x_0 + \delta, z_0)$ and $(x_0 - \delta, z_0)$ within the inner region, respectively, and $\delta$ is a constant. Atoms in the FCC lattice, within SF, or belonging to dislocations are identified via common neighbor analysis[40] implemented in OVITO[41].

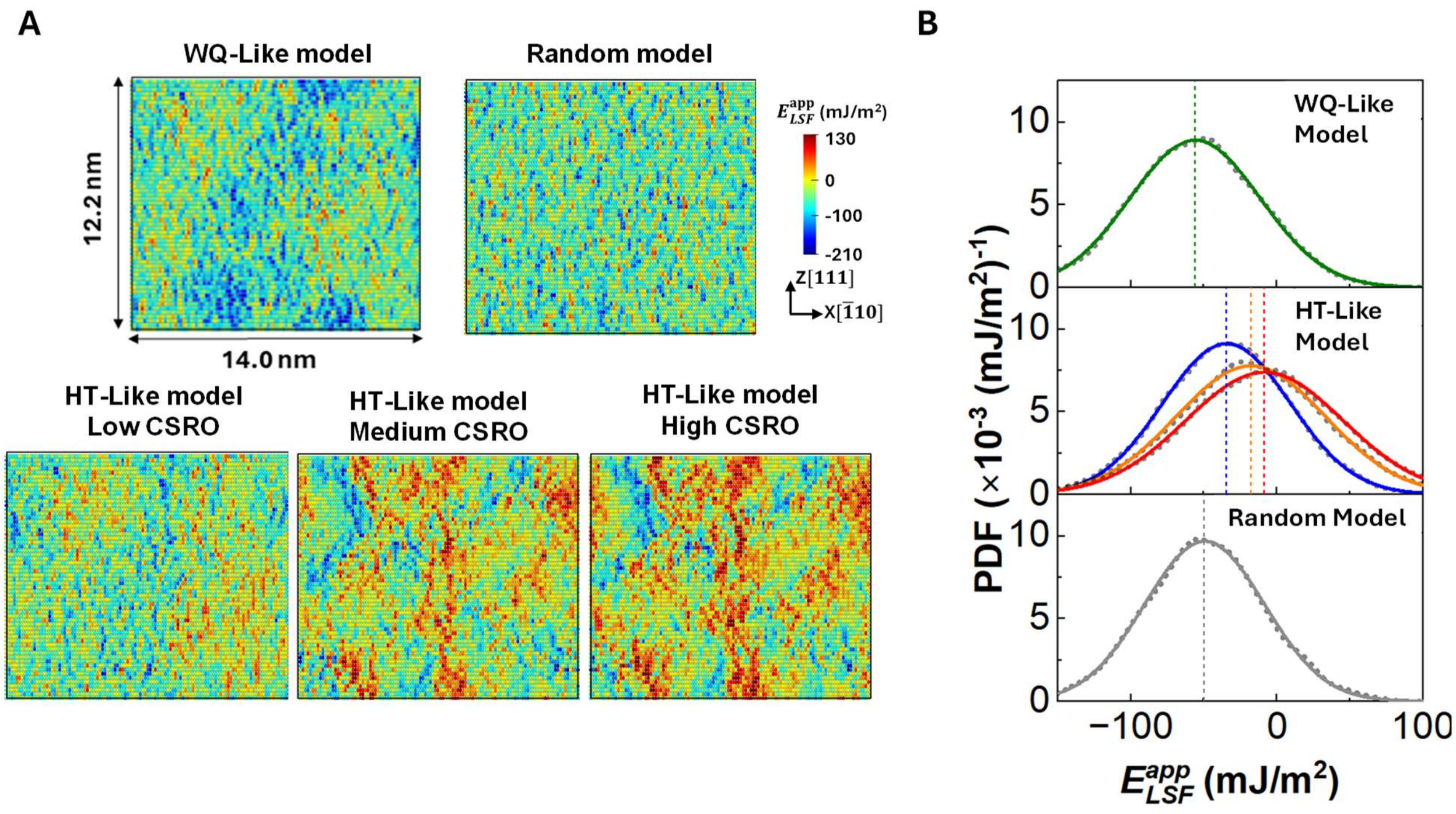

**Fig. 4 Simulated apparent local stacking-fault energetics.** A) Spatial distributions of the calculated $E_{\mathrm{LSF}}^{\mathrm{app}}$ in the WQ-like, random and HT-like CrCoNi models over a 112 × 60 grid. The HT-like models

represent different realizations of $L1_2$-type local ordering. B) PDFs of $E_{\mathrm{LSF}}^{\mathrm{app}}$ for the same models. The random model is shown in gray as a reference without CSRO, while the WQ-like model (olive) represents a distinct CSRO state. In the HT-like models, three distributions corresponding to low, medium, and high degrees of CSRO are shown in blue, orange, and red, respectively. Dotted curves denote the Gaussian-kernel density estimates of the PDFs with the bandwidth selected according to Silverman's rule of thumb[42]. Solid curves represent Gaussian fits based on Eqn. 2, and the vertical dashed lines indicate the fitted mean $E_{\mathrm{SF}}$ values.

Fig. 4A illustrates the heterogeneous spatial distribution of the calculated $E_{\mathrm{LSF}}^{\mathrm{app}}$ in the WQ-like, random, and HT-like CrCoNi models. Among these three models, the random model exhibits the most homogeneous distribution, whereas the highly ordered HT-like model displays the greatest spatial heterogeneity. The PDFs of the calculated $E_{\mathrm{LSF}}^{\mathrm{app}}$ are reasonably described by Gaussian fits (Fig. 4B), and the fitted parameters are summarized in Table 1. The HT-like medium- and high-CSRO models shift the $E_{\mathrm{SF}}$ upward relative to the random model, whereas the WQ-like model remains lower. Specifically, the fitted $E_{\mathrm{SF}}$ values for the HT-like medium- and high-CSRO models are -17.7 and -8.38 mJ $m^{-2}$, respectively, with corresponding distribution widths of 51.4 and 54.0 mJ $m^{-2}$. These values approach the experimentally inferred near-zero $E_{\mathrm{SF}}$ range.

**Table 1.** The fitting parameter $E_{SF}$ and the standard deviation (σ) of the distribution of $E_{\mathrm{LSF}}^{\mathrm{app}}$ in the WQ-like, HT-like and random CrCoNi atomistic models.

| | WQ-like model | HT-like model | | | Random model |
|---|---|---|---|---|---|
| | | Low CSRO | Medium CSRO | High CSRO | |
| $E_{SF}$ (mJ $m^{-2}$) | −56.0 | −34.4 | -17.7 | -8.38 | −49.5 |
| σ (mJ $m^{-2}$) | 44.8 | 43.8 | 51.4 | 54.0 | 41.2 |

While the simulations here are not intended to provide a one-to-one quantitative reproduction of the experiment, the results do show that experimentally plausible changes in the spatial arrangement and degree of CSRO can shift the apparent local fault-energetics distribution. The remaining differences in the absolute $E_{\mathrm{SF}}$ and distribution width likely reflect the non-unique construction of the atomic models, finite model size and the limited experimental constraints on the real CSRO configuration.

***History-dependent statistical fault energetics***

A distributed SFE landscape affects how dislocations sample local chemical environments during the continued slip. Because dislocation motion can modify the surrounding chemical order, the effective fault energetics can become history dependent. This coupling can bias the competition among the planar slip, twinning and HCP transformation, with the latter two critically dependent on the interplanar interactions. This effect can be investigated by examining the history-dependent generalized SFEs (HD-GSFEs) without requiring that the simulations uniquely reproduce the full experimental microstructure[43]. Along this line, we calculated HD-GSFEs for the WQ-like and HT-like low-CSRO models (Suppl. Note 6; Table. S1) and used calculations to examine how different local-ordering motifs can alter the energetic preference for continuous slip, twinning, and HCP transformation.

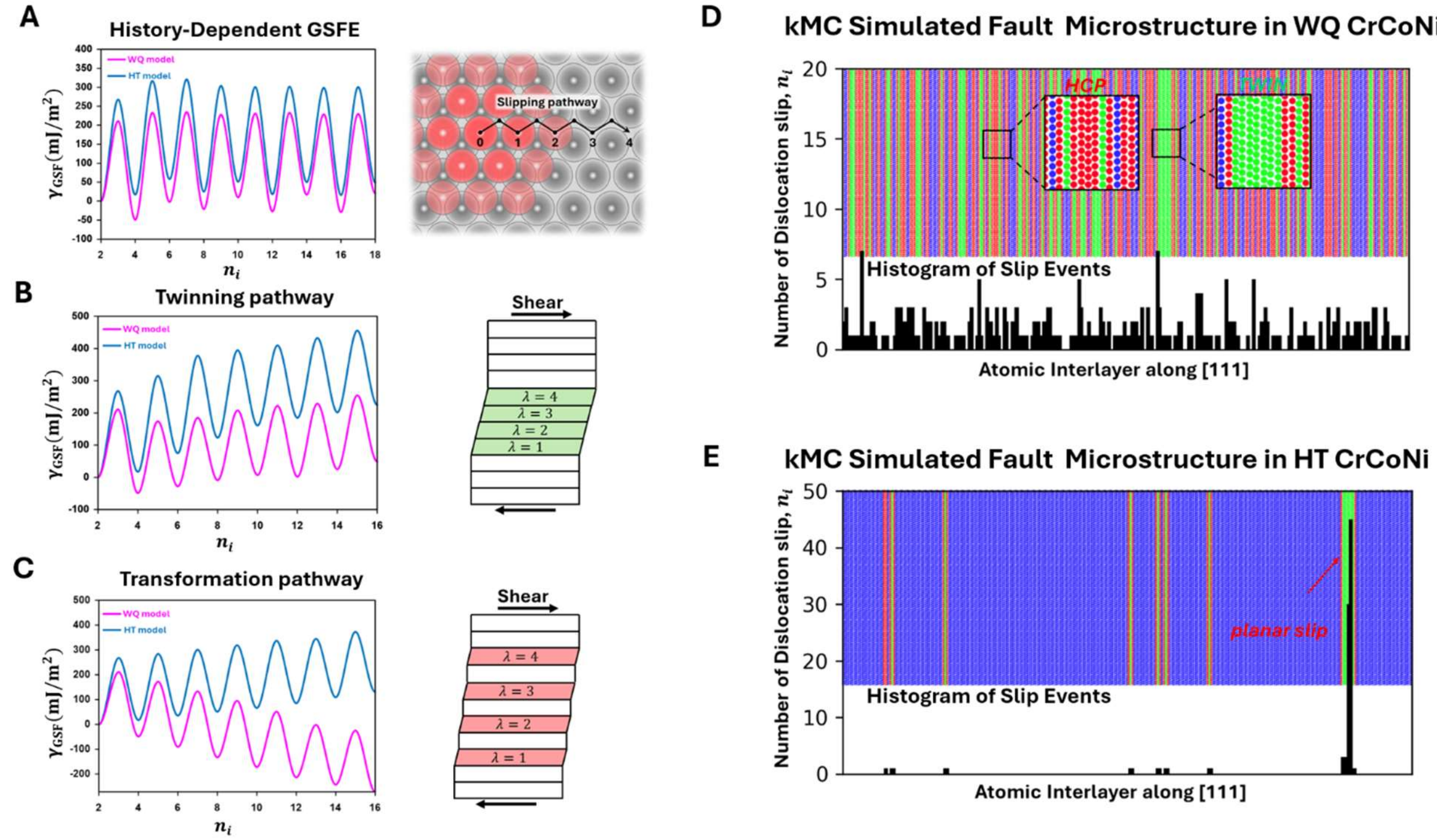

**Fig. 5 Computational modelling of stacking-fault energetics and fault substructures in WQ-like and HT-like CrCoNi models.** A) Example calculated HD-GSFE profiles for the WQ-like and HT-like models. Here, the displacement coordinate is expressed in units of a full Burgers vector, as illustrated schematically. Atoms in the stacking-fault region are shown in red in the ...ABA... stacking sequence. B, C) Examples of calculated twinning and HCP-transformation energy landscapes as functions of λ, the slipping interlayer number in the schematic diagrams. D, E) One-dimensional kMC-simulated fault substructures based on the HD-GSFE event list in Table. S1. Blue, green and red atoms denote FCC matrix, twin and HCP lamellae, respectively. Bar plots show the slip history of each atomic interlayer; high slip counts indicate localized planar slip.

The calculated HD-GSFE for the WQ-like model (Fig. 5A) shows that the initial slip lowers the energy, consistent with a negative apparent SFE, whereas the HT-like low-CSRO model exhibits a positive initial SFE in this simulation. It should be noted that the positive initial GSFE of the selected HT-like configuration in Fig. 5A refers to a specific slip plane and local ordering realization, and thus it should not be identified with the spatially averaged apparent local SFE reported for the same model class in Fig. 4. With the continued slip, the GSFEs do not fully recover, reflecting the disruption of local atomic order caused by dislocation motion. Overall, the $L1_1$-type motif lowers the GSFE profile, whereas the $L1_2$-type motif raises it in the simulated configurations, indicating different tendencies for the SF formation.

The computed energy pathways for twinning and HCP transformation (Figs. 5B, C) indicate that $L1_2$-type ordering can increase the barriers for twin propagation and HCP development, whereas the $L1_1$-type motif reduces these barriers in the WQ-like model. These simulations illustrate how CSRO-dependent HD-GSFE landscapes can bias the competition among the slip, twinning, and HCP transformation.

We then performed 1D kMC simulations to explore how these energetic differences influence fault-substructure evolution (Methods, Suppl. Note 7). The WQ-like model produces a higher density of SFs and broader twin lamellae, whereas the HT-like model accommodates deformation primarily through localized and occasionally intense slip on individual interlayers.

**Experimental evidence for processing-dependent deformation pathways**

The 1D-kMC prediction is supported the ex-situ TEM observation of deformation substructure in the deformed samples . Figure 6 examines the planar faults and their microstructure in the heavily deformed WQ and HT samples using cepstral STEM (Fig. 1, Suppl. Note 9) and atomic-resolution imaging. The color-coded planar fault images in Fig. 6 were collected from three different regions in the deformed WQ and HT samples, respectively. These images show that in the WQ sample, extensive planar faults were formed with well-defined SFs (red), NTs (green) and HCP phases (blue). The NTs range from several to tens of nanometers in thickness. In contrast, the HT sample features more complex and segmented planar faults, often composed of alternating regions of SFs, thin NTs, and short HCP segments within a single planar fault, up to hundreds of nanometers in length. The cepstral STEM results are further supported by atomic-resolution imaging (Figs. 6C-D), which show thin NTs and isolated SFs in the HT sample. The massive nanotwins and extensive HCP phase formation in the WQ sample are consistent with the 1D-kMC prediction (Fig. 5D). In the HT sample, faulted bands are more segmented and often contain alternating SFs, thin NTs, and short HCP lamellae. The mixed fault morphology in the HT sample is a key distinguishing feature and reflects the influence of the dominant CSRO type on deformation pathways.

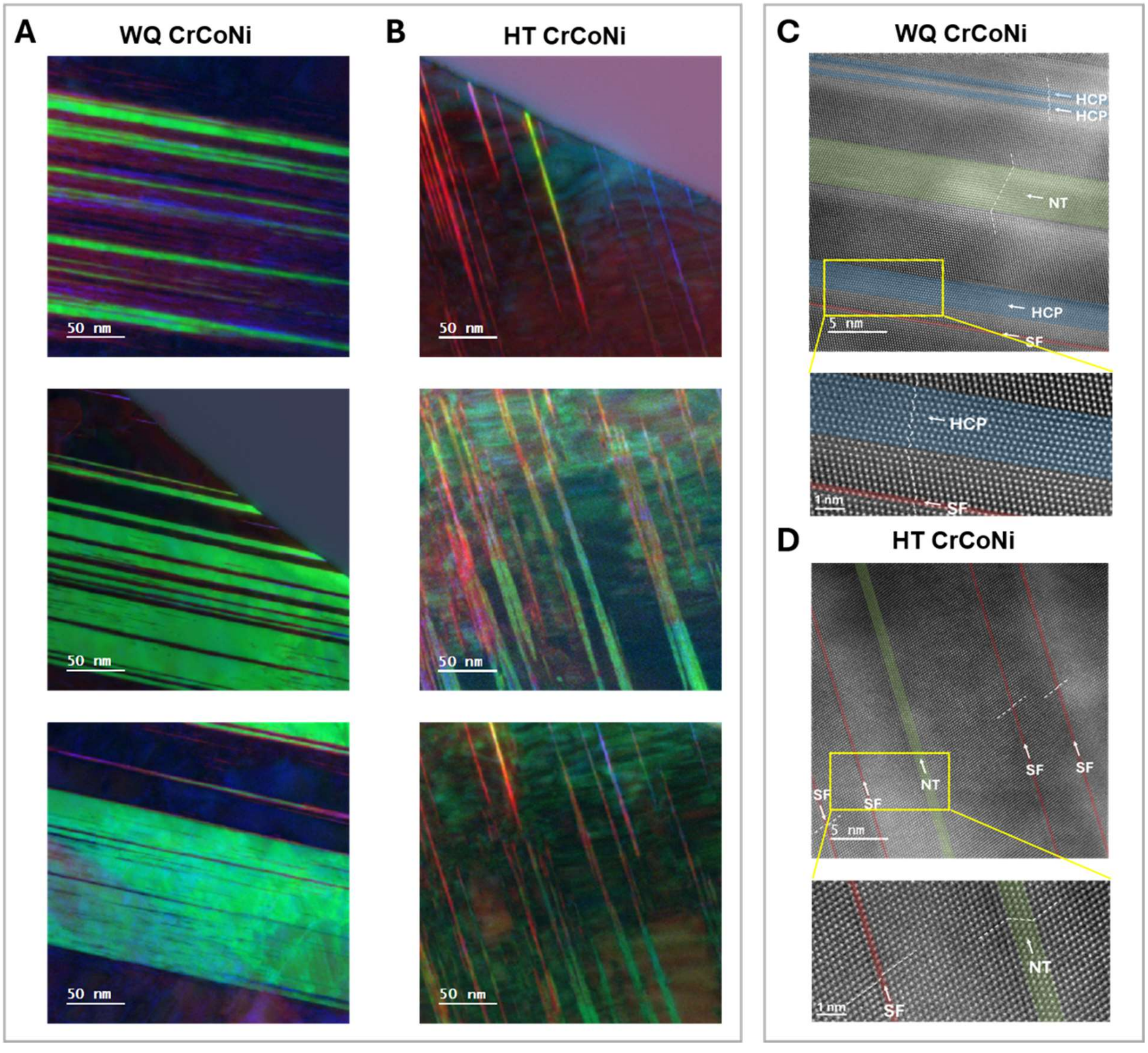

Fig. 6 Fault substructures in tensile-tested WQ and HT CrCoNi samples. A, B) Color-coded cepstral STEM images of three selected regions within a grain in the WQ and HT samples, respectively; SFs, NTs and HCP lamellae are shown in red, green and blue. C, D) Atomic-resolution HAADF-STEM images from heavily deformed regions of the WQ and HT samples, showing the associated atomic arrangements and fault substructures.

Electron imaging (Figs. 6C, D and Figs. S5 and 6) is consistent with the cepstral STEM maps, revealing thin NTs and isolated SFs in the HT sample and more extensive nanotwinning and HCP lamellae in the WQ sample.

## DISCUSSION

### *The most probable stacking fault width, apparent SFE ($\gamma_{\mathrm{app}}$) , and twinning stress*

In CCAs with a distributed apparent fault energetics, the most probable SF width ($L_{\mathrm{m}}$) and its associated probability density ($P_{\mathrm{m}}$) provide an experimentally accessible summary of the underlying energy landscape sampled by dissociated dislocations. In our model, these quantities can be converted into a model-based representative apparent SFE, $\gamma_{\mathrm{app}} \equiv \frac{f(G, b_{\mathrm{p}}, \upsilon, \beta)}{L_{\mathrm{m}}}$, and distribution width ($\sigma$). This yields

$$\gamma_{\mathrm{app}} = E_{\mathrm{SF}} + \frac{\sigma^2}{\gamma_{\mathrm{app}}}, \quad \text{or} \quad E_{\mathrm{SF}} = \gamma_{\mathrm{app}}\left(1 - \frac{\sigma^2}{{\gamma_{\mathrm{app}}}^2}\right) \qquad (6)$$

where $\sigma$ characterizes the width of the underlying apparent local SFE energy distribution. For pure metals, $\sigma = 0$ and $\gamma_{\mathrm{app}} = E_{\mathrm{SF}}$ as expected. For CCAs, equation 6 predicts that $\gamma_{\mathrm{app}} = \left[E_{\mathrm{SF}} + \sqrt{E_{SF}^2 + 4\sigma^2}\right]/2$ when the intrinsic $E_{\mathrm{SF}}$ is negative, thus predicting $\gamma_{\mathrm{app}}$ is determined by the magnitude of fluctuations in local fault energetics in this case.

Comparing this expression with the phenomenological form of Werner[35,44],

$$\gamma_{\mathrm{app}} = E_{\mathrm{SF}} + b_{\mathrm{p}}\tau_{\mathrm{tw}}, \qquad (7)$$

which expresses the experimentally determined apparent SFE $\gamma_{\mathrm{app}}$ as a sum of the intrinsic SFE and the experimentally observed twinning stress, $\tau_{\mathrm{tw}}$, scaled by the Burgers vector. The interpretation is that the $\gamma_{\mathrm{app}}$ is not an intrinsic materials property but depend on the path of partial dislocations under the influence of a resolved shear stress that leads to the stacking fault formation. The addition of $b_{\mathrm{p}}\tau_{\mathrm{tw}}$ to the thermodynamically defined intrinsic $E_{\mathrm{SF}}$ was found to be consistent with experimental observations in Cu-Al alloys and CrMnFeCoNi M/HEAs with low SFEs, where an inverse dependence of twinning stress on SFE emerges once the SFE falls below a critical value[45,46]. In this low-SFE regime, the twinning stress is primarily controlled by twin-band propagation through constrictions of extended forest dislocations[47].

Relating eqn. 6 to 7, one obtains

$$b_{\mathrm{p}}\tau_{\mathrm{tw}} \propto \frac{\sigma^2}{\gamma_{\mathrm{app}}}. \qquad (8)$$

This relation implies $\tau_{\mathrm{tw}} \propto \frac{\sigma^2}{\gamma_{\mathrm{app}}}$, which suggests a statistical interpretation of phenomenological relations between apparent SFE and twinning resistance, arising from fluctuations in the $E_{\mathrm{LSF}}^{\mathrm{app}}$ landscape. This connection, as implicated by our model rather than as a complete microscopic derivation, deserves to be examined further in future work for its underlying mechanism.

***In-situ neutron diffraction evidence for processing-dependent deformation pathways***

In-situ neutron diffraction provides a bulk-sensitive means of monitoring the evolution of deformation pathways during mechanical loading. We performed such experiment during tensile loading to track stacking-fault probability (SFP), dislocation density, and texture evolution in bulk WQ and HT specimens. Neutron diffraction patterns were collected continuously during monotonic loading, so defect-related peak shifts and peak broadening could be followed as functions of engineering strain. This enables deformation mechanisms to be tracked continuously as a function of strain, providing a statistically representative, bulk-sensitive complement to the local, post-mortem observations obtained by electron microscopy.

Figures 7A,B plot the calculated stacking fault probability (SFP) and dislocation density (DD) from the in-situ neutron diffraction data (Suppl. Note 8). The SFP values were calculated from lattice-strain differences between adjacent reflections (for example, {200} and {400}). Dislocation density evolution was quantified using the convolutional multiple whole-profile (CMWP) method[51].The WQ sample exhibits a larger SFP magnitude than the HT sample over most of the measured strain range, indicating a higher diffraction signature associated with stacking faults in the WQ state and lower stacking-fault activity in the HT state. As shown in Fig. 7B, both samples exhibit comparable dislocation densities over the strain range of 5-30%, suggesting broadly similar levels of dislocation accumulation despite differences in faulting activity.

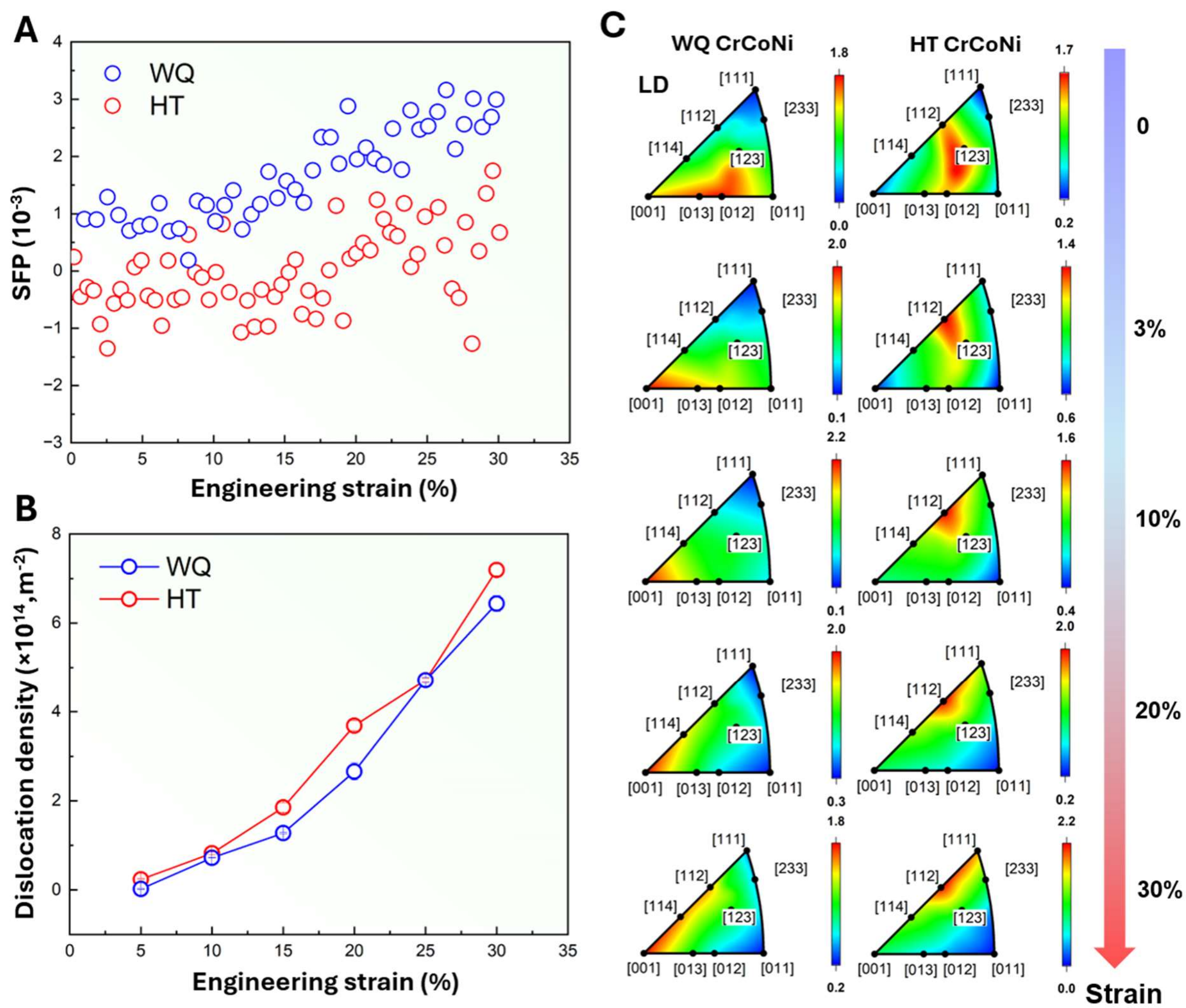

**Fig. 7 In-situ neutron diffraction measurements of water-quenched (WQ) and heat-treated (HT) CrCoNi alloys.** A) SFP of both alloys as a function of engineering strain. The sign follows the diffraction-analysis convention; the comparison of fault activity is based on the magnitude/defined SFP metric. B) Dislocation density as a function of engineering strain. C) Evolution of inverse pole figures (IPFs) along the loading direction (LD) as a function of engineering strain.

Fig. 7C shows the texture evolution of WQ and HT CrCoNi during tensile deformation up to 30% engineering strain. The two tested samples were produced from the same cast and followed the same thermo-mechanical processing route before the final heat treatments (Methods). Prior to deformation, the WQ sample exhibits a strong <012> texture, whereas the HT sample displays a dominant <123> texture. As deformation proceeds, both samples trend toward <112>/<111>-type orientations at higher strains, although their evolution pathways differ (Suppl. Note 8).

The <012> texture in the WQ sample is generally less favorable for the SF/twinning/HCP activity than the <123> texture in the HT sample[34]. Despite this texture tendency, the in-situ neutron diffraction data show a lower SFP magnitude in the HT sample. Thus, the observed SFP difference is not readily explained by the texture alone. Rather it is consistent with an important contribution from differences in the local atomic structure, including CSRO, between the WQ and HT states.

The in-situ neutron diffraction results here demonstrate differences in fault-mediated deformation pathways, which according to our analysis can be attributed to the difference in the distribution of the deformation relevant apparent fault energetics in the HT and WQ state. This result suggests a route for using local chemical order to bias deformation pathways in chemically complex alloys.

## MATERIALS AND METHODS

### Sample preparation

The CrCoNi alloys were prepared from high-purity (>99.9 wt%) Cr, Co, and Ni by arc melting and drop casting under an argon atmosphere, as described in Ref.[8]. Two samples were studied. The WQ sample was homogenized at 1,200 °C for 48 h and then water-quenched. The HT sample was homogenized under the same condition, further aged at 1,000 °C for 120 h and then furnace-cooled to room temperature. The samples were monotonically tensile-tested to fracture at a nominal strain rate of $1 \times 10^{-3}$ $s^{-1}$, reaching a strain of approximately 55%[8]. All post-deformation microstructural characterizations were performed on heavily deformed regions with comparable strain magnitudes.

### STEM characterization and 4D-STEM acquisition

TEM samples were prepared from lightly and heavily deformed regions of the tensile-tested bars by mechanical cutting and polishing, followed by twin-jet electropolishing to electron transparency in an electrolyte of 95 vol.% ethanol and 5 vol.% perchloric acid at -40 °C and 30 V. The grains selected for microscopy were oriented close to the <110> zone axis, with the loading direction in the foil plane. Electron imaging was performed using a probe-corrected Themis Z STEM (Thermo Fisher Scientific) operated at 300 kV. Atomic-resolution HAADF-STEM images were acquired with a semi-convergence angle of 21.4 mrad and a HAADF inner collection angle of 40 mrad. 4D-STEM datasets were acquired over regions of interest using a probe semi-convergence angle of 0.46 mrad and a probe full width at half maximum of 1.7 nm. Diffraction patterns were recorded using an EMPAD detector at 1,000 frames $s^{-1}$.

**Measurements of SF widths using cepstral STEM**

Cepstral STEM[36] was used to identify SFs, NTs, and HCP lamellae and to map their distributions over fields of view hundreds of nanometers wide with approximately 1 nm spatial resolution. The principles and validation of cepstral STEM are described in Suppl. Note 1. To measure the SF width, we first identified isolated SFs in cepstral STEM images collected from multiple regions in the heavily deformed WQ and HT samples and then measured their widths using the procedure described in Suppl. Note 1. In total, 167 and 74 isolated SFs were measured for the WQ and HT samples, respectively. The measured SF widths range from several nanometers to ~260 nm, and their distributions are shown in Fig. 1.

**Statistical inference of representative apparent fault energetics and distribution width from the measured stacking-fault-width distribution**

The experimental SF-width distributions were analyzed using Gaussian KDE and bootstrap resampling (Suppl. Note 4). For each sample, the measured SF widths $L_i$ were used to construct a KDE-based probability density function $p(L)$ with a Gaussian kernel. The kernel bandwidth $(h)$ was selected according to bootstrap-based uncertainty and boundary-sensitivity criteria for $L_\mathrm{m}$, together with a stability criterion based on the derived scaled stacking fault energy $(\widetilde{E_\mathrm{SF}})$. The selected bandwidths were $h_\mathrm{WQ}$ = 4.0 nm and $h_\mathrm{HT}$ = 6.1 nm, respectively, for the WQ and the HT samples. To quantify uncertainty of $L_\mathrm{m}$, we performed 5000 bootstrap resampling iterations by resampling the measured SF widths with replacement and without boundary reflection; for each bootstrap sample, the KDE peak $(L_\mathrm{m}, P_\mathrm{m})$ were recorded to calculate the standard deviation of the $L_\mathrm{m}$ as its uncertainty. The boundary-sensitivity of $L_\mathrm{m}$ to the physical boundary at $L = 0$ was quantified by the fraction of the $L_\mathrm{m}$ less than two bandwidths from $L = 0$ in the bootstrap resampling without boundary reflection, $P(L_\mathrm{m} < 2h)$. As the bandwidth increases, the bootstrap uncertainty of $L_m$decreases, whereas the fraction of realizations affected by the lower boundary generally increases. To evaluate the sensitivity of the derived $\widetilde{E_\mathrm{SF}}$, we solved the $\widetilde{E_\mathrm{SF}}$ as a function of bandwidth using the $L_\mathrm{m}$ and $P_\mathrm{m}$ given by the Gaussian-kernel KDE with the full observed SF widths. The long-width tail was analyzed both with and without truncation at the observed maximum width of ~260 nm. Based on these analyses, the selected bandwidths were $h_\mathrm{WQ}$ = 4.0 nm and $h_\mathrm{HT}$ = 6.1 nm where the standard deviation of the $L_\mathrm{m}$ has decreased to a plateau, the fraction of realizations affected by the lower boundary remains moderate [$P(L_\mathrm{m} < 2h) \approx 0.15$], and the derived $\widetilde{E_\mathrm{SF}}$ exhibits only weak dependence on bandwidth. The resulting $(L_\mathrm{m}, P_\mathrm{m}) = (9.06 \pm 1.29$ nm, $0.0344 \pm 0.0029$ nm$^{-1}$) and ($13.56 \pm 2.86$ nm, $0.0274 \pm 0.0029$ nm$^{-1}$) at the selected bandwidth and with truncation at the observed maximum SF widths for the WQ and HT samples, respectively, are reported as the best estimate with its uncertainty estimated from the standard deviation of bootstrap samples.

These values were then used with the statistical force-balance model in Suppl. Note 4 to infer model-based apparent SFE ranges as a function of dislocation character $\beta$.

**Construction of Atomistic Models with Varying Degrees of CSRO for Simulations**

To model the WQ and HT states, atomistic configurations were constructed following the structural motifs reported by Hsiao et al.[8]: a WQ-like model with dominant $L1_1$-type ordering and HT-like models with dominant $L1_2$-type ordering. The models have dimensions of 14.0 nm × 13.8 nm × 12.2 nm and contain 215,040 atoms (see Fig. 3A). The X, Y and Z axes were aligned along $[\bar{1}10]$, $[\bar{1}\bar{1}2]$ and $[111]$, respectively, with periodic boundary conditions in all three directions. Because the precise atomic arrangements of CSRO in the experimental specimens cannot be uniquely resolved, CSRO was modeled by embedding locally ordered $L1_1$ or $L1_2$ clusters (see Fig. 3B) into a FCC matrix, followed by atomic pair swaps to disrupt long-range periodicity while retaining local ordering. The exact number of swaps and resulting CSRO metrics are provided in Suppl. Note 5. The WQ-like and HT-like models should be interpreted as experimentally motivated realizations based on dominant ordering motifs rather than unique atomistic reconstructions of the measured specimens. To examine the sensitivity of apparent local SFE to CSRO realization, we additionally constructed HT-like models with low, medium and high degrees of $L1_2$-type local order by varying cluster composition, volume fraction, and the number/history of pair swaps.

**Atomistic simulation of apparent local SFE**

The apparent local SFE, $E_{\mathrm{LSF}}^{\mathrm{app}}$, is the derivative of the energy per unit dislocation length ($e$) with respect to the position of the partial, $\mathrm{d}e/\mathrm{d}x$. To calculate the $E_{\mathrm{LSF}}^{\mathrm{app}}$, a cylindrical volume of atoms was extracted from the periodic arrangements of the constructed WQ and HT CrCoNi models. This cylinder, centered at an arbitrary position ($x_0$, $z_0$) with the cylindrical axis aligned along the Y direction, has a radius of 3 nm (see Fig. 3C). To calculate $\mathrm{d}e/\mathrm{d}x$ with respect to the dislocation position via the finite difference method, a single straight partial dislocation with a Burgers vector of 1/6[$\bar{1}\bar{1}2$] was introduced at the positions of ($x_0$−δ, $z_0$) and ($x_0$+δ, $z_0$) together with the isotropic displacement field for the Y direction: $u_y(x, z; x_0 \pm \delta, z_0) = \frac{b_p}{2\pi}\arctan\frac{z-z_0}{x-(x_0\pm\delta)}$, where $b_p$ is the magnitude of the Burgers vector, and δ is a constant to infinitesimally shift the partial. Thus, one SF was positioned at the left side of the partial dislocation and extended to the cylinder surface (Fig. 3C). The cylindrical boundary introduces an image force whenever the dislocation deviates from the center of the cylindrical model at ($x_0$, $z_0$). However, the partial dislocation at ($x_0$−δ, $z_0$) and ($x_0$+δ, $z_0$) experience forces of nearly equal magnitude but opposite direction. The image forces generated by the cylindrical boundary conditions effectively cancel each other out. Consequently, the $E_{\mathrm{LSF}}^{\mathrm{app}}$ at ($x_0$, $z_0$) is given by: $E_{\mathrm{LSF}}^{\mathrm{app}}(x_0, z_0) = E_{\mathrm{LSF}} + f_0 = \frac{e(x_0+\delta,z_0)-e(x_0-\delta,z_0)}{2\delta}$ , where $e(x,z) = E(x,z)/L_y$ is an averaged energy of whole model per unit model thickness, $L_y$ is the length of partial dislocation which is equal to the model thickness in Y and is equal to the typical length of dislocation observed in the experimental samples, and $E(x,z)$ is the energy of the whole model.

The $E(x,z)$ was computed using the conjugate gradient energy minimization method as implemented in LAMMPS[48] with a neural network interatomic potential for CrCoNi[49]. During the energy minimization, the cylindrical model was divided into three regions: inner, intermediate, and outer regions with different constraints. Atoms in the inner region, which is a cylinder with a radius of 0.38 nm and centered at $(x_0 + \delta, z_0)$ or $(x_0 - \delta, z_0)$), were constrained to relax only along the Y direction for a fixed dislocation position. Atoms in the intermediate region, which surrounds the inner region with an outside radius of 2.00 nm and is centered at $(x_0, z_0)$), were unconstrained, while atoms in the outer region were fully fixed. We used $\delta = 0.62$ Å, corresponding to half of the average atomic plane spacing along the X direction, which was sufficiently small to evaluate the $E_{\mathrm{LSF}}^{\mathrm{app}}(x_0, z_0)$. The inclusion of relaxation in the X and Z directions would make the initially straight dislocation line to become slightly wavy, but we expect that the averaged position of the partial dislocation deviates very little from $(x_0 + \delta, z_0)$ or $(x_0 - \delta, z_0)$ compared to the total displacement distance $2\delta$ used in the finite difference method. The main contribution to "local" SFE in the $E_{\mathrm{LSF}}^{\mathrm{app}}(x_0, z_0)$ comes from the increase in the SF area inside the inner region. Validation of this method by calculating $E_{\mathrm{LSF}}^{\mathrm{app}}$ for FCC Ni (123 mJ $m^{-2}$) with the same atomic model showed excellent agreement with the SFE calculated using a slab model (122 mJ $m^{-2}$), taking into account the very small lattice friction force on the partial dislocation in FCC Ni, $f_0 \sim 0$, according to Ref. 56.

**One-dimensional kMC simulation**

Details of the 1D-kMC method are provided in Refs.[43,50]. The method was modified to include HD-GSFE profiles with multilayer coupling effects in $L1_1$- and $L1_2$-type CSRO regions (Suppl. Note 6). The HD-GSFE values were obtained from molecular-statics calculations of constructed fault models (Suppl. Note 5). Slip of an interlayer was treated as a kMC event under the assumption of homogeneous partial-dislocation nucleation. The reaction rate for slip of the $i$-th interlayer was estimated using an Arrhenius expression, with the activation free energy evaluated at temperature T and external shear stress on the (111) plane. The event list for all transition states is provided in Table. S1, and additional details are given in Suppl. Note 6.

***In-situ* neutron diffraction**

In-situ neutron diffraction experiments under tensile loading were performed on the VULCAN Engineering Materials Diffractometer at the Spallation Neutron Source, Oak Ridge National Laboratory, using an MTS load frame[51]. The experiments were conducted at room temperature with an incident beam size of 8 mm × 3 mm. A continuous strain-control mode was used during tension at a strain rate of $6.8 \times 10^{-6}$ $s^{-1}$, measured with an MTS 634.12F-21 static axial clip-on extensometer. Neutron diffraction patterns were collected continuously during monotonic loading and reduced using VDRIVE[52]. The profiles were analyzed for SFP (Suppl. Note 8) and dislocation density using the convolutional multiple whole-profile (CMWP) method[53-55]. In the SFP analysis, positive and negative values arise from the sign convention of the peak-shift model; they do not indicate positive or negative physical faulting. Fault activity is compared using the magnitude or consistently defined SFP metric described in Suppl. Note 8. IPFs were extracted from selected in-situ diffraction patterns using the method in Ref.[56].

## ACKOWLEDGMENTS

The present work was supported by DMR-2226495 from the Metals and Metallic Nanostructures Program (MMN) within the Division of Materials Research (JMZ, PL) and PPG company (to KJY). SO was supported by Japan Society for the Promotion of Science (JSPS) KAKENHI Grant Numbers JP18H05453 and JP23K20037 and Ministry of Education, Culture, Sport, Science and Technology of Japan (MEXT) Programs, Grant Numbers JPMXP1122684766, JPMXP1020230325, and JPMXP1020230327. PY acknowledges the financial support by the National Natural Science Foundation of China (NSFC) (Grant no. 52301010) and Hainan University (Grant no. RZ2300002798). We used resources at the Spallation Neutron Source, a DOE Office of Science User Facility operated by the Oak Ridge National Laboratory. The beam time was allocated to VULCAN on proposal number IPTS-25375.1. The authors thank Dr. Dunji Yu for his help with neutron-diffraction measurements. Electron microscopy experiments were conducted in the Materials Research Laboratory Central Research Facilities, University of Illinois. The calculations were performed on the large-scale computer systems at the D3 Center, The University of Osaka, the Large-scale parallel computing server at the Center for Computational Materials Science, Institute for Materials Research, Tohoku University, Research Center for Computational Science, Okazaki, Japan (Project: 25-IMS-C503), and supercomputer Fugaku provided by the RIKEN Center for Computational Science (Project IDs: hp250229 and hp250227). The authors acknowledge the use of ChatGPT (GPT-5.6 Luna, OpenAI) for language editing during manuscript preparation.

## AUTHOR CONTRIBUTIONS

JMZ, PL, and SO directed the project. JMZ and KJY developed cepstral STEM for imaging deformation faults. KJY performed fault microstructure analysis and stacking fault width measurements. RF synthesized the samples and performed in-situ neutron diffraction analysis together with KA. JPD developed the statistical model of stacking fault width distribution and performed atomistic simulations. PJY simulated the deformation substructure and calculated the history dependent stacking fault energies. All authors contributed to the writing of the manuscript.

### Data availability

The source data, including atomic models and simulation outputs, will be deposited in (insert repository name and DOI/accession link) upon publication.

### Code availability

The codes supporting this study's findings are publicly available from sources cited in relevant references.

# Supplementary Materials

## Evidence for Distributed Fault Energetics and Their Impact on Deformation in a Chemically Complex Alloy

Kaijun Yin[1†], Jun-Ping Du[2†], Peijun Yu[2,3†], Rui Feng[4†], Hanyu Hou[1,5], Haw-Wen Hsiao[1], Ke An[4], Peter K. Liaw[6], Shigenobu Ogata[2*], and Jian-Min Zuo[7,8,1*]

[1]Department of Materials Science and Engineering and Materials Research Laboratory, The Grainger College of Engineering, University of Illinois at Urbana-Champaign, Urbana, IL, 61801, USA

[2]Department of Mechanical Science and Bioengineering, The University of Osaka, Osaka, 560-8531, Japan

[3]State Key Laboratory of Materials for Advanced Nuclear Energy, Shanghai University, Shanghai, 200444, China

[4]Neutron Scattering Division, Oak Ridge National Laboratory, Oak Ridge, TN, 37831, USA

[5]Center for Nanoscale Materials, Argonne National Laboratory, Lemont, IL, 60439, USA

[6]Department of Materials Science and Engineering, University of Tennessee, Knoxville, Knoxville, TN, 37996, USA

[7]Monash Center for Electron Microscopy, Monash University, VIC, Australia

[8]Department of Materials Science and Engineering, Monash University, VIC, Australia

[†]These authors contributed equally to this paper

[*]Corresponding authors: jianzuo@illinois.edu, ogata@me.es.osaka-u.ac.jp

**Contents:**

- Supplemental Notes 1 to 9
- Supplemental Tables 1 to 2
- Supplemental Figures 1 to 7

## Supplementary Notes

### *Supplementary Note 1. Cepstral STEM imaging and identification of SFs, NTs, and HCP phases*

The principle of cepstral STEM for planar defect determination is illustrated in Fig. 1 of main text. First, nanodiffraction datasets are collected over regions of interest at every probe position using the setup for 4D-STEM and using a nanobeam. The collected diffraction datasets can either be used to form virtual dark-field (VDF) STEM images or analyzed using the difference cepstrum method developed by *Shao et al.*[36] for cepstral VDF imaging (Fig. 1B). The latter analysis involves calculating the difference between two cepstral transforms of a local nanodiffraction pattern, $[I(\vec{k})]$, and the averaged nanodiffraction pattern over a reference region, $[I_{avg}(\vec{k})]$, which is free of defects. The resulting quantity, $dC_p$, is determined by the following formula[36]:

$$dC_p = \left|FT\left\{log\left[\frac{I(\vec{k})}{I_{avg}(\vec{k})}\right]\right\}\right| = \left|FT\{log[I(\vec{k})]\} - FT\{log[I_{avg}(\vec{k})]\}\right|, \qquad \text{(S1)}$$

For the interpretation of cepstral STEM images, we consider the $dC_p$ analysis of zone-axis diffraction patterns. First, there is a correspondence between the exit wave and the projected atomic structure, even in case of strong dynamical scattering, as first suggested by Geuens and Van Dyck[57]. By expressing the exit wave function as a sum of exit waves from individual atomic columns, we obtain

$$\varphi_e(\vec{r}) = \sum_i \varphi_i(\vec{r} - \vec{r}_i), \qquad \text{(S2)}$$

where $i$ indexes the atomic columns under the nanobeam and $\vec{r}_i$ is the position of the atomic column, and the Fourier transform of the exit wave gives

$$\varphi_e(\vec{k}) = \sum_i \varphi_i(\vec{k}) e^{2\pi i \vec{k}\cdot\vec{r}_i}. \qquad \text{(S3)}$$

Under the above approximation, the diffraction intensity is given by

$$I(\vec{k}) = \varphi_e \varphi_e^* = \sum_i \sum_j \varphi_i(\vec{h}) \varphi_j^*(\vec{k}) e^{2\pi i \vec{k}\cdot(\vec{r}_i - \vec{r}_j)}, \qquad \text{(S4)}$$

If we consider the individual atomic column exit wave functions similar, the Fourier transform of the diffraction intensity then gives the following autocorrelation function:

$$FT[I(\vec{k})] = \sum_i \sum_j FT\left(\left|\varphi_i(\vec{k})\right|^2\right) * \delta(\vec{r} - \vec{r}_i + \vec{r}_j), \qquad \text{(S5)}$$

which is the same as the Patterson function, with the pseudo atomic-scattering factor, $\varphi_i(\vec{k})$. The effect of the *log* function in the cepstral transform is a scaling of diffraction intensities, which sharpens the pseudo atomic-density distribution and thus produces narrower correlation peaks. According to Eq. (S5), the measured $dC_p$ pattern can be interpreted as the difference between a local Patterson function and the averaged one.

Fig. S1 shows the $dC_p$ patterns obtained from three different diffraction patterns (DPs) with diffuse streaks from SFs and extra diffraction spots from NTs and HCP phases, respectively. In all three $dC_p$ patterns, harmonic peaks parallel to the corresponding fault or interface plane are observed. The features are weak in the case of Fig. S1A for the SF, indicating a small number of contributing atomic pairs within the fault. In comparison, harmonic peaks parallel to the corresponding interfaces are strong for NTs or HCP lamellae with increase in the number of faulted layers. The $dC_p$ patterns in Fig. S1B and C further reveal additional harmonic peaks from NTs and HCP phases, which are specific to each fault-related structure. These harmonic peaks can be associated with the arrangements of atoms and their interatomic vectors as illustrated in the atomic models of Fig. S1. Hence, both the intensity and harmonic peaks in $dC_p$ patterns are highly sensitive to the nature of the fault-related structures, and thus cepstral VDF imaging provides unique signatures for the analysis of fault-related structures.

Using the harmonic peak signals marked by circles in Fig. S1, we generated three types of cepstral STEM images from a 4D-STEM dataset acquired from the WQ sample, highlighting HCP phases, NTs, and SFs (Fig. 1C). The distribution of these fault-related structures is made clearer in the color-coded image, where blue, green, and red correspond to HCP lamellae, NTs, and SFs, respectively. In this color-coded image, regions with strong red signals indicate the presence of isolated SFs where HCP and NT signals are weak or absent.

The ability of cepstral STEM to distinguish SFs, NTs, and HCP lamellae is further validated by atomic-resolution imaging, as shown in Fig. S2, in which two SFs separated by 2.6 nm are imaged by both cepstral and annular-dark-field (ADF) STEM. In both cases, the SFs are clearly detected and resolved. The spatial resolution of cepstral STEM is approximately 1 nm (corresponding to the pixel size), while ADF-STEM resolves individual atomic columns, respectively. In Fig. S2C and D, mixed fault-related structures comprising HCP lamellae, NTs, and isolated SFs are observed. Regions exhibiting a combination of these features are indicated by mixed colors of blue (HCP lamellae), green (NT) and red (SF), and are shown in greater detail in the ADF-STEM image.

These results demonstrate that cepstral STEM offers a significantly larger field of view and the ability to analyze slightly off-zone-axis orientations and relatively thick samples and provides complementary information to atomic-resolution imaging.

***Supplementary Note 2. Stacking fault characters***

The SFs observed in the WQ and HT samples are of mixed types, as determined by the $\beta$ angle between the perfect dislocation Burgers vector and the dislocation line. The determination of $\beta$ angle can be made through atomic resolution imaging, when the SFs are in thin regions of the sample. Fig. S3 shows atomic-resolution analysis of two selected SFs, observed in the WQ sample. One is a screw dislocation with $\beta$ = 0°, bordered by 30° Shockley partials on both sides (Figs. S3A). Another is a mixed dislocation with $\beta$ = 60°, where the left partial is a 90° Shockley type, and the right partial is a 30° Shockley type.

***Supplementary Note 3.*** **Apparent local stacking fault energy distribution**

The probability density function (PDF) of the distribution of $E_{\mathrm{LSF}}$ in the FCC M/HEAs with or without CSRO at different dislocation positions with random position sampling can be expressed as a Gaussian distribution[20,38,58,59], $\sim N(E_{\mathrm{SF}}, \sigma_{\mathrm{LSF}}^2)$. Here, the $E_{\mathrm{SF}}$ is the average value of the $E_{\mathrm{LSF}}$. Since the $E_{\mathrm{LSF}}$ is estimated using the sampling at different dislocation positions, their average, $E_{\mathrm{SF}}$, is equivalent to the system's average SFE. The variance of $E_{\mathrm{LSF}}$ is denoted by $\sigma_{\mathrm{LSF}}^2$. While the probability density function of $f_0$ can be also reasonably assumed as a Gaussian distribution[60,61], $\sim N(0,\sigma_{f_0}^2)$, with a mean of zero and a variance $\sigma_{f_0}^2$, resulting from the net force per unit dislocation length —averaged along the dislocation line — exerted by numerous surrounding solute atoms whose individual forces vary randomly in both magnitude and direction. The $E_{\mathrm{LSF}}$ and the $f_0$ can be regarded as approximately independent because they originate from distinct mechanisms. Specifically, the $f_0$ results from the misfit volumes of solutes located above and below the slip plane[62], whereas the $E_{\mathrm{LSF}}$ is caused by the local structural transformation of atoms on the slip plane from a FCC to a hexagonal close-packed (HCP) arrangement forming the SF[37]. Then, the PDF of $E_{\mathrm{LSF}}^{\mathrm{app}} = E_{\mathrm{LSF}} + f_0$ also follows a Gaussian distribution, $\sim N(E_{\mathrm{SF}}, \sigma^2)$, with a mean of $E_{\mathrm{SF}}$ and a variance of $\sigma^2 = \sigma_{\mathrm{LSF}}^2 + \sigma_{f_0}^2$, as expressed by $g\left(E_{\mathrm{LSF}}^{\mathrm{app}}\right)$ in Equation (2) of the main text.

**Supplementary Note 4. Determination of representative stacking fault energy from experimental measurements**

**Kernel density estimation of the stacking fault width distribution:** The PDF of the SF width was estimated using kernel density estimation (KDE) with a Gaussian kernel[42]. The experimentally observed SF widths were truncated at the maximum measurable width $L_{\max} \approx 260$ nm (defined by the field of view of the 4D-STEM mapping),and the truncated KDE was defined as

$$p^{tr}(L) = \frac{1}{Nh}\frac{1}{\sqrt{2\pi}}\sum_{i=1}^{N} \exp\left(-\frac{(L-L_i)^2}{2h^2}\right) \qquad \text{(S6)}$$

where $L_i$ are the individual SF widths, $N = 167$ and $N = 74$ are the total numbers of SF widths measured for the WQ and HT samples, respectively, and $h$ is the bandwidth.

Since negative SF widths ($L < 0$) are physically inaccessible and our analysis focuses exclusively on the peak of the PDF, the nonzero Gaussian KDE density in the $L < 0$ region arising from the infinite support of the Gaussian kernel was neglected. This contribution was verified to have a negligible effect on the estimated peak position.

The effect of truncation at $L_{\max}$was accounted for by defining the corresponding untruncated PDF as

$$p^{\mathrm{untr}}(L) = p^{\mathrm{tr}}(L)\, F(L_{\max}), \qquad \text{(S7)}$$

where $F(L) = \int_0^L p(L')\, dL'$is the cumulative distribution function, given by

$$F(L) = \frac{1-\mathrm{erf}\left(\frac{\frac{1}{L}-\widetilde{E_{\mathrm{SF}}}}{\tilde{\sigma}}\right)}{1+\mathrm{erf}\left(\frac{\widetilde{E_{\mathrm{SF}}}}{\tilde{\sigma}}\right)}. \qquad \text{(S8)}$$

The final peak position and height, $(L_{\mathrm{m}}, P_{\mathrm{m}}) = (L_{\mathrm{m}}, P_{\mathrm{m}}^{\mathrm{tr}})$, were obtained from the truncated KDE constructed using the full SF width dataset at the selected bandwidth. The scaled stacking fault energy $\tilde{E}_{\mathrm{SF}}$ and the scaled $\tilde{\sigma}$ were then determined experimentally from the measured values of $(L_{\mathrm{m}}, P_{\mathrm{m}}^{\mathrm{tr}})$ by solving the two equations: $L_m = \frac{2}{\widetilde{E_{\mathrm{SF}}}+\sqrt{\widetilde{E_{\mathrm{SF}}}^2+4\tilde{\sigma}^2}}$ and $P_{\mathrm{m}}^{\mathrm{tr}} = p(L_m)$. For the untruncated KDE, the same equations were solved, except that the second equation was replaced by $P_{\mathrm{m}}^{\mathrm{untr}} = P_{\mathrm{m}}^{\mathrm{tr}} F(L_{\mathrm{max}}) = p(L_{\mathrm{m}})$. Finally, the model-based apparent SFE ($E_{\mathrm{SF}}$) as a function of dislocation character β was given by the relationship $\widetilde{E_{\mathrm{SF}}} = \frac{E_{\mathrm{SF}}}{f(G,b_{\mathrm{p}},\nu,\beta)}$, where $f\left(G, b_{\mathrm{p}}\nu, \beta\right) = \frac{Gb_{\mathrm{p}}^2}{8\pi}\left(\frac{2-\nu}{1-\nu}\right)\left(1 - \frac{2\nu\cos(2\beta)}{2-\nu}\right)$ is a function of the shear modulus $G$, the Burgers vector magnitude $b_{\mathrm{p}}$ of the partials, Poisson's ratio $\nu$, and the angle $\beta$ between the full Burgers vector and the dislocation line.

**Bootstrap assessment of statistical uncertainty**: Bootstrap resampling was employed to quantify the statistical uncertainty and robustness of the estimated peak position[63]. For each bandwidth $h$, bootstrap samples of the same size as the original SF width dataset were generated by sampling with replacement. For each bootstrap realization, the density peak position $L_m$ was identified within a predefined interval (~0.01 nm). The number of bootstrap realizations was tested using $B = 3000$ and $B = 5000$, and no statistically significant differences were observed; therefore, $B = 5000$ was adopted throughout.

**Bandwidth selection and boundary sensitivity analysis**: The optimal bandwidth for the Gaussian KDE was selected by jointly considering (i) statistical stability of the peak position, (ii) sensitivity to the lower boundary at $L = 0$, and (iii) robustness of the derived scaled stacking fault energy $\tilde{E}_{\mathrm{SF}}$.

For each bandwidth, the bootstrap distribution of $L_m$was obtained, and the standard deviation of $L_m$was used as a measure of statistical uncertainty. Boundary sensitivity was quantified by the fraction of bootstrap realizations for which the peak position lies within two bandwidths of the lower boundary, $P(L_m < 2h)$, which reflects the effective support of the Gaussian kernel and the potential influence of boundary-induced bias.

As the bandwidth increases, the bootstrap uncertainty of $L_m$decreases, whereas the fraction of realizations affected by the lower boundary generally increases (Figs. S4A and B). The bandwidth was therefore selected from the plateau region of the $L_m$ uncertainty as a function of bandwidth, where further increases in $h$ lead to only marginal reductions in statistical uncertainty, while avoiding the regime in which boundary sensitivity becomes dominant.

In addition, we required that the derived scaled stacking fault energy varies only weakly with bandwidth in the selected range (Figs. S4C and 4D), ensuring that the reported physical results are not driven by over smoothing.

As an additional diagnostic, we verified that the fraction of bootstrap realizations for which $L_m < h$ remains below 1% at all tested bandwidths, $P(L_m < h) < 0.01$, indicating that strong boundary-induced bias is negligible.

Applying these criteria, the bandwidth was chosen as $h = 4.0$ nm for the WQ dataset and $h = 6.1$ nm for the HT dataset. These values correspond to regions where the bootstrap uncertainty of the peak position has reached a plateau, the fraction of realizations affected by the lower boundary remains moderate, and the scaled stacking fault energy exhibits only weak dependence on bandwidth.

***Supplementary Note 5. Settings in the atomistic model construction***

The construction of the atomistic models is guided by two key experimental observations[8]:

(i) In the WQ sample, the predominant CSRO is of the $L1_1$-type at the nanometer scale, whereas in the HT sample, the dominant CSRO is the $L1_2$-type. Notably, a significant portion of the system remains chemically disordered, and the local elemental compositions within both $L1_1$ and $L1_2$ CSRO nanoclusters deviate from equiatomic compositions.

(ii) The overall elemental composition of the system is equiatomic.

For the construction of the WQ-like CrCoNi model, perfect $L1_1$-type nanoclusters with a composition of Cr(CoNi), in which Cr atoms occupy the ordered close-packed planes and Co and Ni equally occupy the remaining sites, chemically random nanoclusters with a composition of $Cr_{25}Co_{37.5}Ni_{37.5}$, and chemically random matrix regions with an equiatomic composition of $Cr_{33.3}Co_{33.3}Ni_{33.3}$ were embedded in the atomistic model with volume fractions of 1/6, 1/3, and 1/2 respectively, ensuring equiatomic compositions of Cr, Co, and Ni. The $L1_1$-Cr(CoNi) and the chemically random $Cr_{25}Co_{37.5}Ni_{37.5}$ nanoclusters were assumed to be spherical with a diameter of ~2.8 nm and were randomly distributed throughout the system. Four possible orientations of the close-packed Cr planes in the $L1_1$ structures (Fig. 3B, main text) were randomly assigned to the $L1_1$-Cr(CoNi) nanoclusters clusters, and overlap between ordered clusters was prohibited, whereas the overlap between chemically random nanoclusters was permitted because no chemical ordering is present within these nanoclusters. To introduce the $L1_1$-type CSRO, swaps between randomly selected unlike atomic pairs were performed within both the $L1_1$-Cr(CoNi) nanoclusters and chemically random $Cr_{25}Co_{37.5}Ni_{37.5}$ nanoclusters. Swapping was continued until the Warren–Cowley parameter[18] $\alpha_{\mathrm{CrCr}}^{m}$ in the $L1_1$ regions was reduced by approximately one-half for neighbor shells with m < 22, which was achieved after 15,000 swaps. The Warren–Cowley parameter is defined as $\alpha_{ij}^{m} = 1 - \frac{p_{ij}^{m}}{c_j}$, where $p_{ij}^{m}$ is the probability of finding an atom of element *j* in the *m*-th neighbor shell around an atom of element *i*, and $c_j$ is the atomic fraction of element *j* in the analyzed region.

For the construction of the HT-like models, perfect $L1_2$-type nanoclusters with a composition of $CrCo_3$, chemically random nanoclusters with a composition of $Cr_{75}Co_{25}$, and pure Ni matrix regions were embedded in the atomistic model with equal volume fractions of 1/3, thereby preserving the overall equiatomic composition. The two types of nanoclusters were spherical with diameters of ~2.8 nm and randomly distributed throughout the system. In contrast to the $L1_1$ structure, the $L1_2$ structure possesses only a single Cr–Cr bond orientation along ⟨100⟩ and were allowed to overlap. Subsequently, 10,000 swaps were performed between Co in $L1_2$-$CrCo_3$ nanoclusters and Ni atoms, and between Cr atoms in the chemically random $Cr_{75}Co_{25}$ nanocluster and Ni atoms to homogenize the chemical distribution. An additional 6,000 swaps between randomly selected unlike atomic pairs were performed to generate the first HT-like model. Continuing the random-swap stage until a cumulative total of 15,000 swaps had been performed in this stage, excluding the preceding 10,000 homogenizing swaps, generated the second HT-like model. To evaluate the sensitivity of the apparent local SFE to different CSRO realizations, the third HT-like model with varying degrees of ordering was constructed by modifying the cluster compositions, volume fractions, and numbers of atomic swaps. In the construction of the third HT-like model, perfect $L1_2$-type nanoclusters with a composition of $Cr(CoNi)_3$, in which Cr atoms occupy the corner sublattice and Co and Ni equally occupy the face-centered sites, chemically random nanoclusters with a composition of $Cr_{50}Co_{25}Ni_{25}$, and chemically random matrix regions with a composition of $Cr_{33.3}Co_{33.3}Ni_{33.3}$ were embedded in the atomistic model with the volume fractions of 1/3, 1/6, and 1/2, respectively, which also fulfill the overall equiatomic composition condition. Subsequently, 15,000 swaps between randomly selected unlike atomic pairs were carried out within both the $L1_2$-$Cr(CoNi)_3$ and chemically random $Cr_{50}Co_{25}Ni_{25}$ nanoclusters to generate the third HT-like model. Based on their first-neighbor Warren–Cowley parameters[18] ($\alpha_{ij}^{1}$), the first, the second, and the third HT-like models were labeled as the low-, medium-, and high-CSRO HT models, respectively, as listed in Table. S1.

***Supplementary Note 6.*** **History-dependent GSFEs calculations**

The SFs were created by rigidly shifting the upper slab of the (111) atomic layers by $\frac{1}{6}a_0(11\bar{2})$. Twinning was generated by the consecutive slip of neighboring layers, while the HCP-phase transformation was modeled by sequentially inserting stacking faults while skipping interlayers, following methods described in *Ref.* [43,49]. The surface-boundary condition is applied to the $\langle 111 \rangle$ direction while the periodic-boundary conditions (PBCs) are applied to the other two directions. To avoid the surface effect, all fault-related structures are placed in the middle of the supercell (215,040 atoms). Due to the random distribution of $L1_1$ and $L1_2$ chemically ordered regions, the SF positions were fixed in the model center, and mean values were taken across interlayers by shifting the entire atomic slab along $\langle 111 \rangle$. The SFE can be calculated as:

$$E_s = (E_f - E_0)/A, \quad \text{(S9)}$$

where $E_f$ is the total energy of the model after creating the SF, $E_0$ is the energy of the perfect crystal after energy minimization, and $A$ is the cross-sectional area of the (111) plane of the model.

The previous work by Yu et al. showed that in the CrCoNi with CSRO, history-dependency (HD) should be considered when introducing fault-related structures into the model[43]. In the present work, the HD-GSFE effect is examined for both the WQ and HT models with the dominant $L1_1$- or $L1_2$-type CSRO regions, respectively. As shown in Fig.5A of main text, the first SF introduced into the WQ model lowered the energy, indicating a negative SFE, whereas the partial $L1_2$ ordering in the HT model maintained a positive SFE. It should be mentioned that in the fully $L1_1$ and $L1_2$ ordered structures calculations, the single GSFE of each case is -190 mJ/m$^2$ and 42 mJ/m$^2$, respectively. When the subsequent $b_p$ is introduced to the same interlayer and the SF recovered to FCC, the energy of the sample is not recovered to the original state. With increasing $n_i$, the GSFEs for both models converge to certain values, which are similar to the behavior of a random solid-solution state CrCoNi studied in the previous work.

***Supplementary Note 7. Kinetic Monte Carlo (KMC) simulation***

In the kMC modeling, each slipping event was modelled as the partial Burgers vector $b_p$ slipping of an interlayer $i$ ($0 \leq i \leq N_0$), assuming homogeneous partial-dislocation nucleation. Thus, a model with $N_0$ interlayers has $N_0$ possible events at each kMC step, with the activation free energy corresponding to that of the homogeneous partial-dislocation nucleation. It is worth noting that partial-dislocation nucleation often occurs heterogeneously in reality, such as at grain boundaries or at defect sites within the grain, resulting in a lower activation energy. For this reason, the modeling of homogeneous nucleation provides a lower bound estimate for the event rate. The rate $\upsilon_i$ of the $i$-th interlayer slipping was estimated using a rate equation, $\upsilon = \upsilon_0 exp(-\Delta G(T,\tau)/k_B T)$, where $\Delta G(T,\sigma)$ is the activation free energy of the partial dislocation nucleation at temperature $T$ and at external shear stress $\tau$ on the $(111)\langle 110\rangle$ direction, $k_B$ is the Boltzmann constant, and $\upsilon_0$ is a prefactor of the event trial frequency.

The activation energy of partial-dislocation loop nucleation is expressed as

$$\Delta G_{m,n}(T,\sigma) = a^c \Delta g_{m,n}(T) + L^c \Gamma_{LT} - a^c b \tau^{ex}, \qquad \text{(S10)}$$

where $\Delta g_{m,n}(T) = \Delta g_{m,n}(0)(1 - \frac{T}{T_m})$. $\Delta g_{m,n}(0)$ is the energy difference at $T = 0$ K and stress-free, which is directly obtained from atomistic simulations and listed in Table. S2. Here, $(1 - \frac{T}{T_m})$ is an approximation of the temperature effect on the activation-free energy. $a^c = \pi r_c^2$ is the activation area where $r_c$ is the nucleation radius as we use a homogeneous partial-dislocation nucleation model with assuming a circular shaped loop nucleus, and $L^c = 2\pi r_c$ indicates the length of the dislocation loop at the saddle point critical radius, $r_c$. $\Gamma_{LT} = \alpha\mu b^2$ represents the dislocation-line tension where $\alpha = 0.123$ is a dimensionless-fitting parameter in the work of Curtin *et al.*[64], $\mu = 187\ GPa$, and $b = 1.457$ Å represent the elastic constant and Burgers vector, respectively. $\tau^{ex}$ indicates the external shear stress. Letting $\Delta f = \pi r_c^2 \Delta g_{m,n}(T) + 2\pi r_c \Gamma_{LT} - \pi r_c^2 b \tau^{ex}$, then $r_c$ is determined by letting $\frac{\partial \Delta f}{\partial r_c} = 2\pi r_c \Delta g_{m,n}(T) + 2\pi \Gamma_{LT} - 2\pi r_c b \tau^{ex} = 0$. The radius of the activation area is then derived as $r_c = \frac{\Gamma_{LT}}{b\tau^{ex} - \Delta g_{m,n}(T)}$. $T_m = 1{,}450\ K$ represents the lattice-surface disordering temperature (the melting temperature) determined by two-phase simulations in our previous work[43]. The deformation temperature used in the kMC simulations is $T = 300\ K$. The transition energy for each transition state can be defined as $\Delta E_k(T,\sigma) = A(\Delta e_k(T) - b\tau^{ex})$, where $A$ is the faults area. Here we set $A = 1\ \mu m^2$ of a typical grain size in such alloys. The temperature dependency on the stable-energy is modeled as $\Delta e_k(T) =$ $\Delta e_k(0)(1-\frac{T}{T_m})$, where $\Delta e_k(0)$ is the energy change for each transition state at 0 K and stress-free, which is directly obtained from the event list that can be seen in Table. S2. We note that the reversed transition (e.g., ISF to FCC or TWIN to HCP) between each state is also based on the nucleation theory. In the current work, the reverse transition refers to further transition caused by the introduction of SFs in the interlayers with the slipping history that would lead to the local structure transit back to its previous state. The models are then strained up to 200% during the kMC simulations.

To mimic the strain rate $\varepsilon = 1 \times 10^{-3} s^{-1}$ in the tensile tests, strain-rate-dependent external shear stress applied during the kMC simulations is derived. Based on the Arrhenius's Law, the total strain rate $\dot{\varepsilon}$ can be written as a function of activation energy

$$\dot{\varepsilon} = \sum_i^N \dot{\varepsilon}_{m,n0} \exp\left(\frac{-\Delta G_{m,n}(T,\tau^{ex})}{k_B T}\right), \qquad \text{(S11)}$$

where $N = 150$ equals to the number of interlayers considered in the model and $\Delta G_{m,n}$ stands for the different activation energies, the prefactor $\dot{\varepsilon}_{m,n0}$ can be approximated as

$$\dot{\varepsilon}_{m,n0} = v_0 n_{m,n} \varepsilon_0 \approx v_0 \frac{D}{d}\left(\frac{b}{D}\right) = v_0 \frac{b}{d}, \qquad \text{(S12)}$$

where $v_0 = 10^{12}\ s^{-1}$ is the attempt frequency of each dislocation nucleation, such as each atomic layer sliding event. $\varepsilon_0$ is the strain generated on a grain by each sliding event. $b$ and $d$ are the Burgers vector of a partial dislocation and the lattice spacing from the MD simulation, respectively. $n_{m,n}$ and $D$ are the number of nucleation sites and average grain diameter, respectively. Here, the event dependent number of nucleation sites $n_{m,n}$ is not only event-dependent, but also time-dependent, thus these change during the deformation process.

Letting $\sum_{m,n}^{N} \dot{\varepsilon_{m,n0}} \exp\left(\frac{-\Delta G \quad ,n(T,\tau^{ex})}{k_B T}\right) = \dot{\varepsilon_0} \exp\left(\frac{-\overline{\Delta G}(T,\tau^{ex})}{k_B T}\right)$, where the $\overline{\Delta G}$ is considered as the apparent activation barrier and Eq. (S12) is used for the apparent prefactor $\dot{\varepsilon_0}$. Then

$$\overline{\Delta G} = -\ln\left(\sum_{i}^{N} \exp\left(\frac{-\Delta G m,n(T,\tau^{ex})}{k_B T}\right)\right) k_B T, \quad \text{(S13)}$$

If we assume the strain rate as $10^{-3}\ s^{-1}$, the $\overline{\Delta G}$ can be determined by inverting the effective Arrhenius relation preceding Eq. (S13)

$$\overline{\Delta G} = -\ln\left(\frac{\dot{\varepsilon}}{\dot{\varepsilon_0}}\right) k_B T, \quad \text{(S14)}$$

Therefore, the strain-rate-dependent external shear stress could be determined by incorporating the formulation of $r_c$ into Eqs. (S13) and (S14). The values of the apparent activation barriers for both RSS and CSRO cases can be computed ($\overline{\Delta G}_{RSS} = 36.92\ eV$ and $\overline{\Delta G}_{CSRO} = 39.11\ eV$), and the corresponding external shear stress can be determined. In the formulation of the apparent activation barrier, the logarithm is highly dependent on the summation of activation frequencies of each transition in the model. Note that these values are only for the initial states of the kMC simulations. The event list changes as the stacking pattern changes during the deformation. Hence, the value of the summation slightly fluctuates depending on the event accepted at each simulation timestep. So are the $\overline{\Delta G}$ and the $\tau^{ex}$. It should also be noted that as the slipping history effect was introduced to the GSFE, it further affects this summation of the activation frequencies of each transition. However, compared to the summarized value, changes in $\tau^{ex}$ made by the HD effect are negligible at the simulation temperature of $T = 300\ K$.

Figs. 5D and E of main text represent the pattern of atomic-plane arrangements in CrCoNi with the $L1_1$ and $L1_2$ CSRO motifs for the simulated WQ and HT samples after a 200% shear strain, respectively. The WQ sample exhibits a large number of SFs due to its negative SFE along with wider twin lamellae than HCP lamellae. Transformations among SFs, twins, and HCP structures were permitted through the HD-GSFE framework in Table. S2, enabling the growth of thicker laminated substructures in the WQ case. In contrast, the HT sample contained only the minimal HCP or twin lamellae, due to the large energy penalties associated with their formation. Instead, strain was mainly accommodated by severe slipping within individual atomic interlayers. This behavior demonstrates that in CrCoNi, $L1_2$ CSRO induces strong history-dependent effects that favor planar slip and inhibit fault-mediated transformations, while $L1_1$ CSRO promotes transformation-driven plasticity.

***Supplementary Note 8. Neutron diffraction analysis***

The *hkl* plane-specific lattice strain of each respective phase is determined by $\varepsilon_{hk} = (d_{hk} - d_{hkl}^{0})/d_{hkl}^{0}$, where $d_{hk}^{0}$ and $d_{hkl}$ are the reference lattice *d*-spacings in the stress-free state and the lattice *d*-spacing during loading, respectively. The stacking-fault probability (SFP) is calculated from lattice strain by the following equation provided by Warren[65]:

$$\varepsilon_{hkl} = \frac{-\sqrt{3}\alpha}{4\pi h_0^2(u+b)} \sum_b(\pm L_0), \quad \text{(S15)}$$

where $\varepsilon_{hkl}$ is the experimentally measured lattice strain for the specific *hkl* reflection, $\alpha$ is the SFP, $h_0^2 = h^2 + k^2 + l^2$ and $L_0 = h + k + l$ for the cubic system, $b$ are the components, i.e., $L_0 = 3N \pm 1$, which are broadened by faulting, and $u$ are the components, i.e., $L_0 = 3N$, which are not broadened by faulting. $N$ is an arbitrary integer. We measured the lattice strains of $\varepsilon_{200}$ and $\varepsilon_{400}$ since the 200 and 400 reflections have higher intensities than those of 111 and 222 in the WQ sample (Fig. S5). For $\varepsilon_{\{200\}}$ and $\varepsilon_{\{400\}}$, the values of $\frac{\sum_b(\pm L_0)}{h_0^2(u+b)}$ are -1/2 and 1/4 for the {200} and {400} reflections, respectively. Equation (S15) can be simplified as follows:

$$SFP = -\frac{16\pi}{3\sqrt{3}}(\varepsilon_{400} - \varepsilon_{200}). \quad \text{(S16)}$$

The relationship between the lattice strain and the applied stress in the elastic regime is used to obtain the elastic constants by the Kroner model (e.g., shear modulus = 80 GPa).

The texture evolution pathway during tensile deformation differed significantly between the WQ and HT samples. While both samples eventually developed strong ⟨112⟩ and ⟨111⟩ textures at high strains, only the WQ sample exhibited a clear intermediate rotation to ⟨001⟩ at 3% and 10% strain before progressing toward ⟨114⟩, ⟨112⟩, and ultimately ⟨111⟩. In contrast, the HT sample transitioned more directly from its initial ⟨123⟩ orientation to ⟨112⟩ and ⟨111⟩, without passing through the ⟨001⟩ orientation. This distinction arises because the initial ⟨123⟩ orientation is crystallographically closer to ⟨111⟩ than ⟨012⟩ (near ⟨001⟩), enabling a more straightforward rotation path under tension. Despite conventional expectations that grains oriented near ⟨111⟩ would promote deformation twinning and stacking fault formation, our observations reveal increased stacking fault activity in the WQ sample. This finding highlights that differences in stacking fault propensity are not governed solely by initial texture or microstructure but rather are strongly influenced by the underlying atomic structure and the dominant types of CSRO.

***Supplementary Note 9. TEM Characterization of the WQ and HT CrCoNi samples***

The CrCoNi alloys were prepared from high purity (> 99.9% weight percent) Cr, Co, and Ni by arc melting and then drop casting under an argon atmosphere as described in Materials and Methods. Two samples were studied here, one was immediately cooled by water quench after being homogenized at 1,200 °C for 48 h (WQ sample), and another was further aged at 1,000 °C for 120 h, followed by furnace cooling to room temperature (HT sample). These two samples were subjected to the monotonic tension test at a strain rate of $1 \times 10^{-3}$ $s^{-1}$ up to the fracture point. The transmission electron microscopy (TEM) samples were prepared from the slightly and heavily deformed areas of the tested-tensile bars, first by mechanical cutting and followed by polishing. The extracted TEM samples were then thinned down to electron transparency by twin-jet chemical etching, using an electrolyte consisting of 95% (volume percent) ethanol and 5% perchloric acid at the temperature of - 40 °C and applied voltage of 30 V.

The slightly deformed CrCoNi alloy samples were previously examined by Hsiao *et al.*[8] for the effect of chemical short-range order (CSRO). Two types of CSRO were reported in the slightly deformed samples and in regions free of dislocation. One type of CSRO minimizes the Cr–Cr nearest neighbors with negative and positive SRO parameters for Cr-(Co/Ni) and Cr-Cr respectively, which is consistent with the $L1_2$-type CSRO. Another type CSRO prefers Cr on alternating close-packed planes, which was identified as the $L1_1$-type CSRO. The CSRO effects are only observed in clusters of a few nm in sizes. The distribution of CSRO nanoclusters changes from being heterogeneous to more homogeneous in WQ and HT sample, respectively. The $L1_1$-type CSRO nanoclusters predominantly form in the WQ sample, while an increase in the number of the $L1_2$-type CSRO nanoclusters and a reduction in both the size and number of the $L1_1$-type CSRO nanoclusters were observed in the HT sample.

Figures. S6 and S7 compare the slightly and heavily deformed samples using bright-field (BF) TEM. The images taken from the slightly deformed WQ and HT samples show more pronounced fault-related structures in the heat-treated sample. In the samples taken from the heavily deformed areas, the representative TEM images show extensive fault-related structures in the WQ sample and arrays of dislocations in HT sample.

## Supplementary Tables

**Table S1.** First-neighbor Warren-Cowley order parameters of the HT-like models with low, medium, and high degrees of CSRO. The parameters and the low-, medium-, and high-CSRO labels refer to the entire atomistic models, not only the $L1_2$-type nanoclusters.

| *HT model* | $\alpha^1_{CrCr}$ | $\alpha^1_{CrCo}$ | $\alpha^1_{CrNi}$ | $\alpha^1_{CoCo}$ | $\alpha^1_{CoNi}$ | $\alpha^1_{NiNi}$ |
|---|---|---|---|---|---|---|
| *High-CSRO* | *-0.29* | *-0.14* | *0.44* | *-0.31* | *0.47* | *-0.93* |
| *Medium-CSRO* | *-0.23* | *-0.11* | *0.34* | *-0.24* | *0.36* | *-0.73* |
| *Low-CSRO* | *0.05* | *-0.02* | *-0.03* | *0.02* | *0.01* | *0.01* |

**Table S2.** Event list of all possible transition states of HD-GSFEs for the WQ and HT CrCoNi models used in this study.

| **Before** | $n_i$ | **GSFE (mJ/m²)** | | **After** |
|---|---|---|---|---|
| | | **HT CrCoNi Model** | **WQ CrCoNi Model** | |
| FCC | 0 | 17 | -49 | ISF |
| | 1 | -23 | -19 | |
| | 2 | -26 | -17 | |
| ISF | 0 | 41 | 37 | FCC |
| | 1 | 16 | 31 | |
| | 2 | 15 | 29 | |
| ISF | 0 | 18 | -42 | HCP |
| | 1 | 1 | -39 | |
| | 2 | 3 | -36 | |
| ISF | 0 | 28 | 21 | ESF |
| | 1 | 9 | 5 | |
| | 2 | 11 | 10 | |
| ESF | 0 | 18 | 19 | TWIN |
| | 1 | 3 | 6 | |
| | 2 | -1 | 4 | |
| TWIN | 0 | 35 | 29 | HCP |
| | 1 | 21 | 14 | |
| | 2 | 22 | 14 | |
| HCP | 0 | 88 | 82 | TWIN |
| | 1 | 59 | 66 | |
| | 2 | 55 | 61 | |

## Supplementary Figures

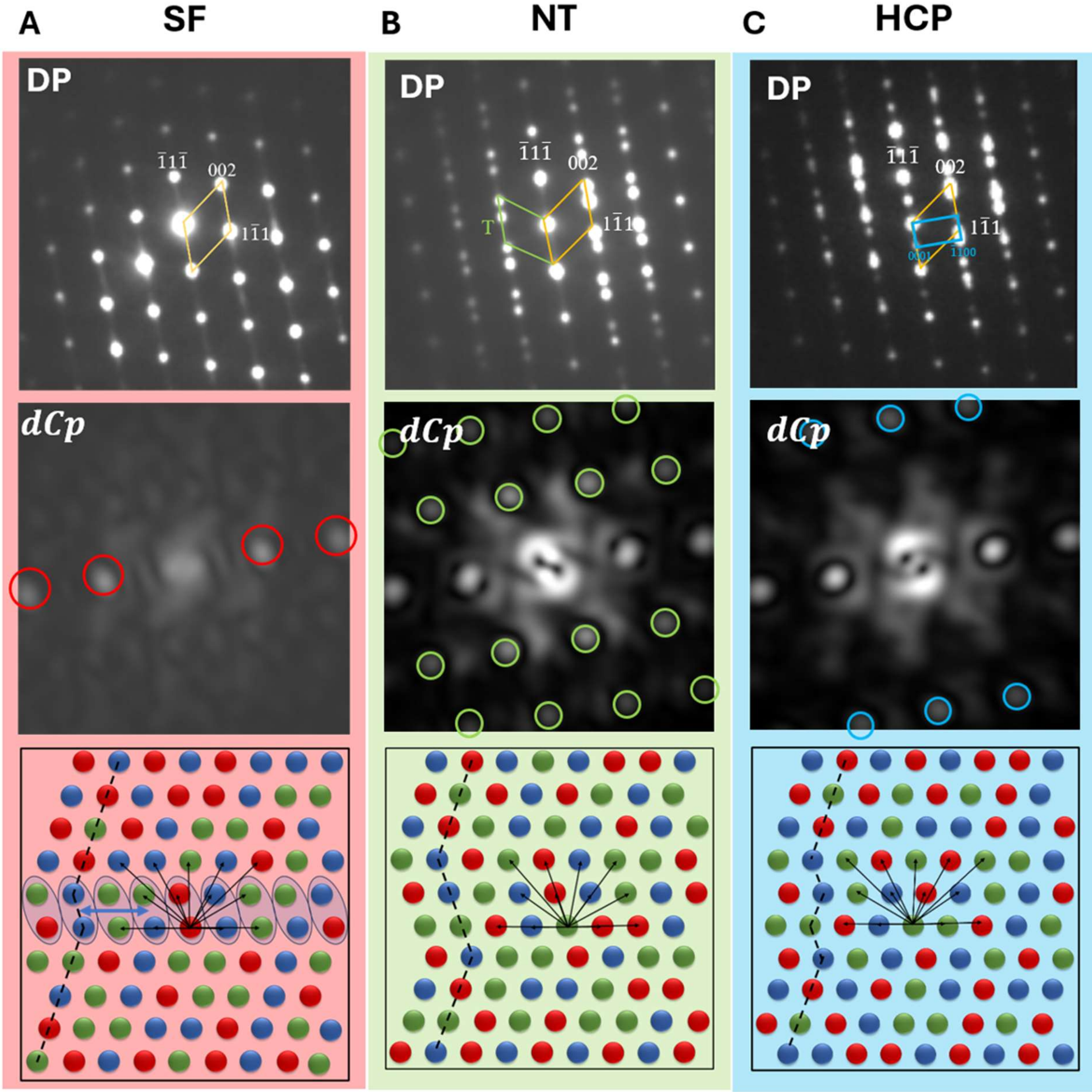

**Fig. S1 Difference cepstral transform ($dC_p$) and relationship to a local atomic structure.** A) (Top) a selected experimental diffraction pattern from a SF in the deformed WQ CrCoNi sample recorded along the [110] zone axis direction, (middle) the corresponding $dC_p$ pattern of the diffraction pattern with notable harmonic peaks marked by the red circles and their positions marked by the yellow arrows, and (bottom) the atomic model of a SF projected along [110] with the ellipses marking the atoms inside the SF and the yellow arrows marking the interatomic vectors, which correspond to the vectors detected in the $dC_p$ pattern. B-C) Same as (A) for a NT and an HCP phase. The yellow arrows mark interatomic vectors in the fault, while the grey arrows mark interatomic vectors in the FCC metal. The yellow and grey arrows overlapped in the faulted region to highlight their difference. The same interatomic vectors in the fault are detected in the $dC_p$ pattern. For comparison, the DPs and the $dC_p$ patterns are displayed at the same intensity scale, separately.

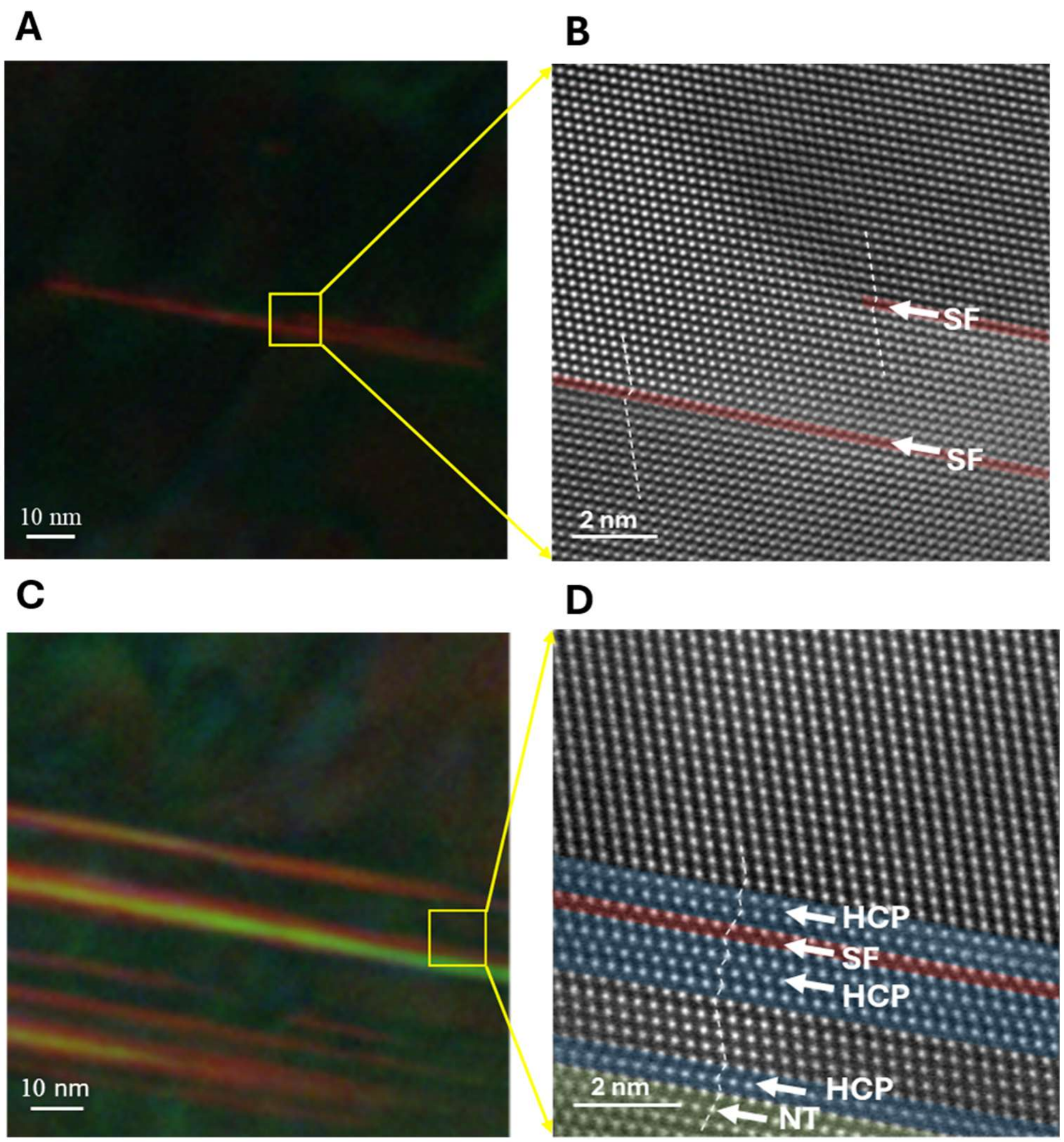

**Fig. S2 Co-located cepstral and atomic-resolution STEM imaging of planar faults in a heavily deformed WQ CrCoNi sample.** B) Atomic-resolution HAADF-STEM images acquired from the regions marked in the cepstral STEM color-coded images in (A), with stacking fault in red. C) Atomic-resolution

HAADF STEM images recorded from regions marked in cepstral STEM color-coded images of (left) with red for SFs, green for NTs and blue for HCP phase. The few-layers HCP phase mixed with a SF in (C) show up in (D) as a mixed color band, because of the resolution limit of cepstral STEM (~1 nm).

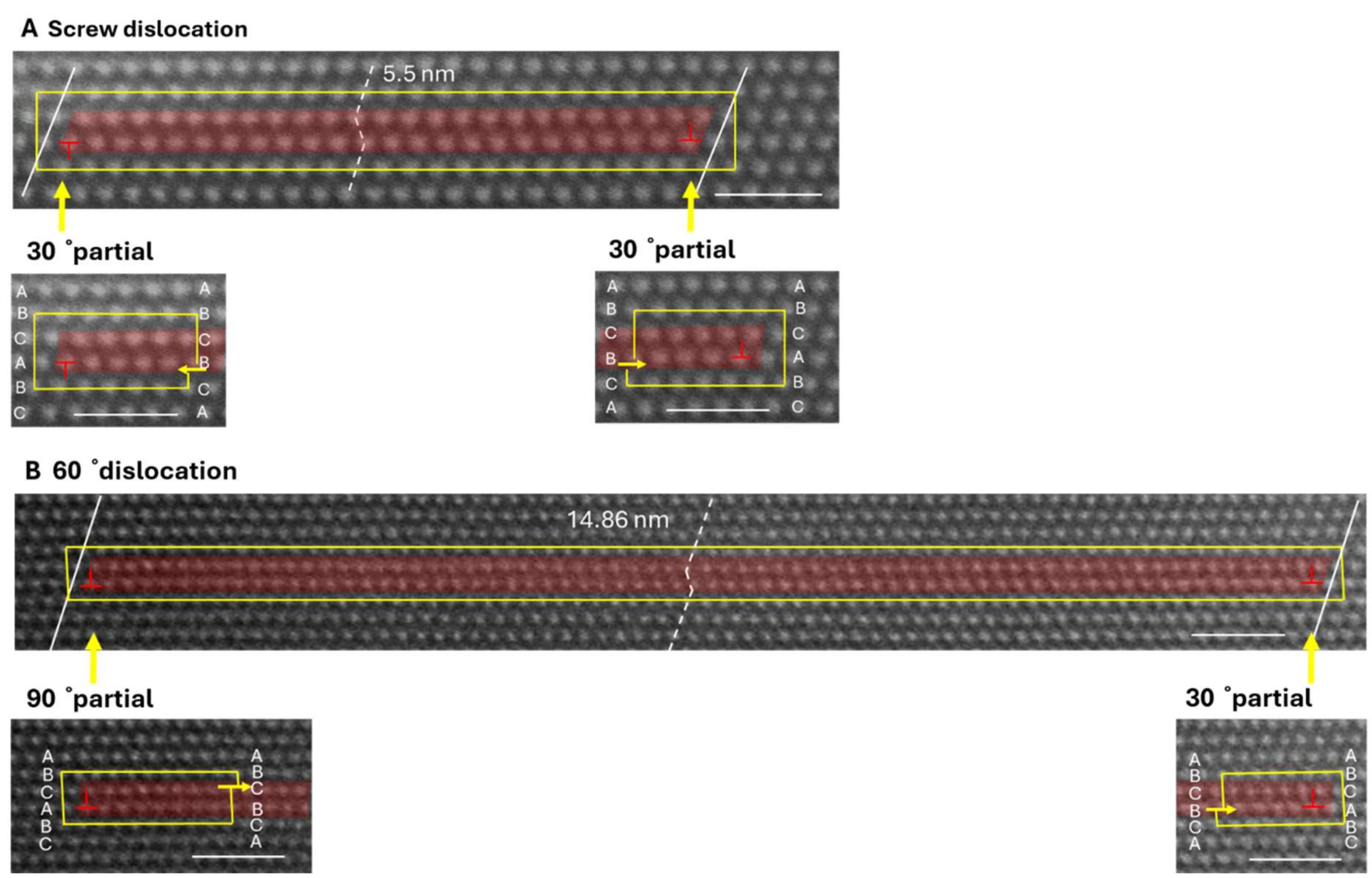

**Fig. S3 Two examples of atomically resolved stacking faults in deformed CrCoNi.** A) A HAADF-STEM image of a screw dislocation with a stacking fault of approximately 5.5 nm wide, terminated by 30° partials. B) A HAADF-STEM image of a ½<110>{111} 60° dislocation with a stacking fault of approximately 14.86 nm, terminated by a 90° and a 30° partials on the left and right, respectively. The small yellow arrows indicate the Burgers-vector direction. Note that the magnification is different for the two dislocations, and the scale bar is 1 nm. Symbol ⊥ represents dislocation.

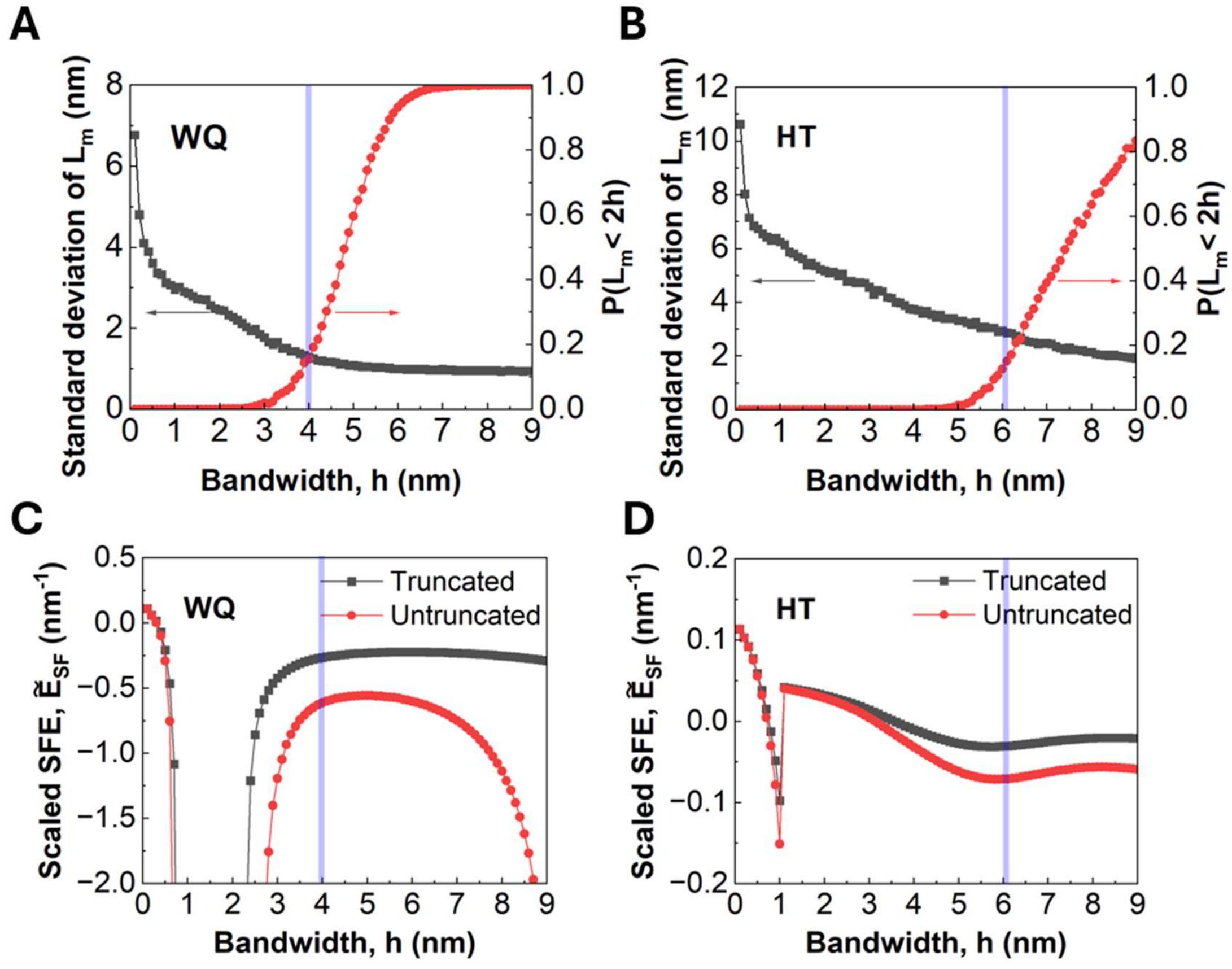

**Fig. S4 Bootstrap stability and bandwidth selection for kernel density estimation of the SF-width probability distribution.** A-B) Bootstrap standard deviation of the peak position $L_{\mathrm{m}}$and the fraction (P) of bootstrap samples with $L_{\mathrm{m}} < 2h$ for the WQ and HT samples. C-D) Scaled stacking fault energy obtained from the KDE constructed using the full SF width dataset as a function of bandwidth for the WQ and HT samples. The shaded region in (A-D) indicates the selected optimal bandwidth, where the standard deviation of the $L_{\mathrm{m}}$ has decreased to a plateau, the fraction of realizations affected by the lower boundary remains moderate, and the scaled stacking fault energy exhibits only weak dependence on bandwidth.

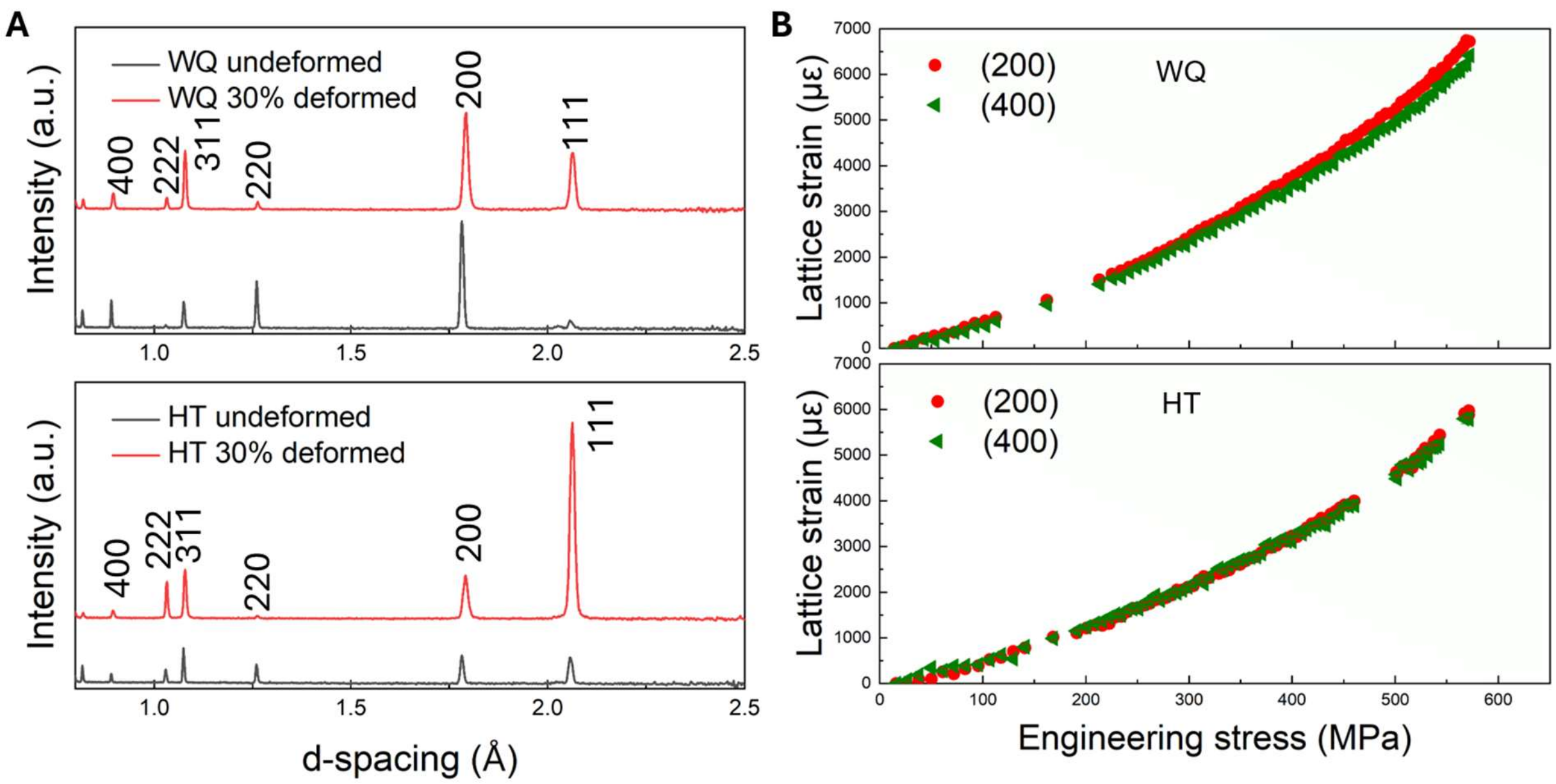

**Fig. S5 Neutron diffraction results of WQ and HT CrCoNi alloys**. A) Neutron diffraction patterns of WQ and HT CrCoNi alloys in their undeformed and 30% deformed conditions. B) Lattice strain responses of the {200} and {400} grains in the WQ and HT CrCoNi alloys as a function of engineering stress.

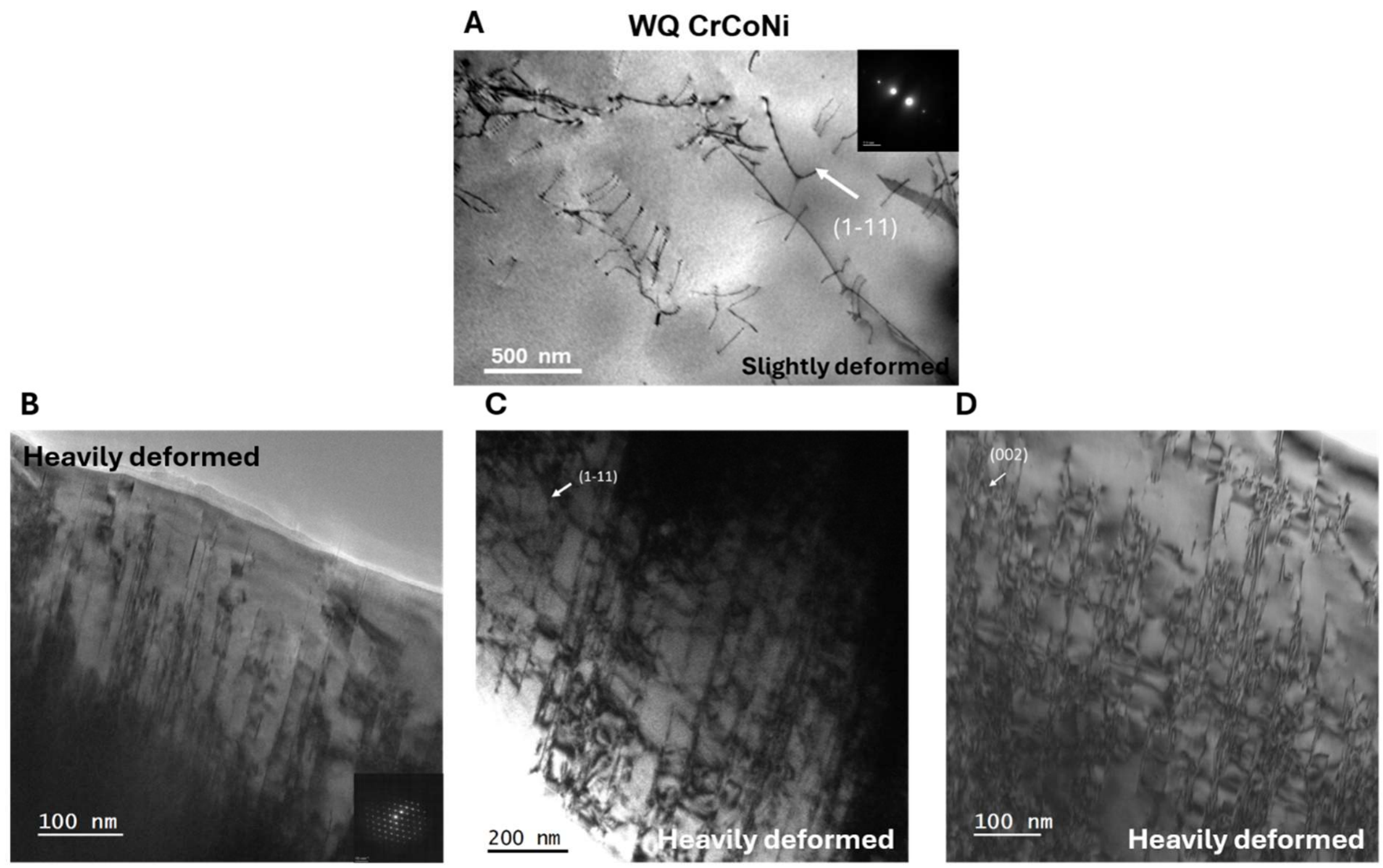

**Fig. S6 TEM analysis of WQ CrCoNi samples.** A) Slightly deformed region of the tensile bar. B-D) Images from the heavily deformed regions.

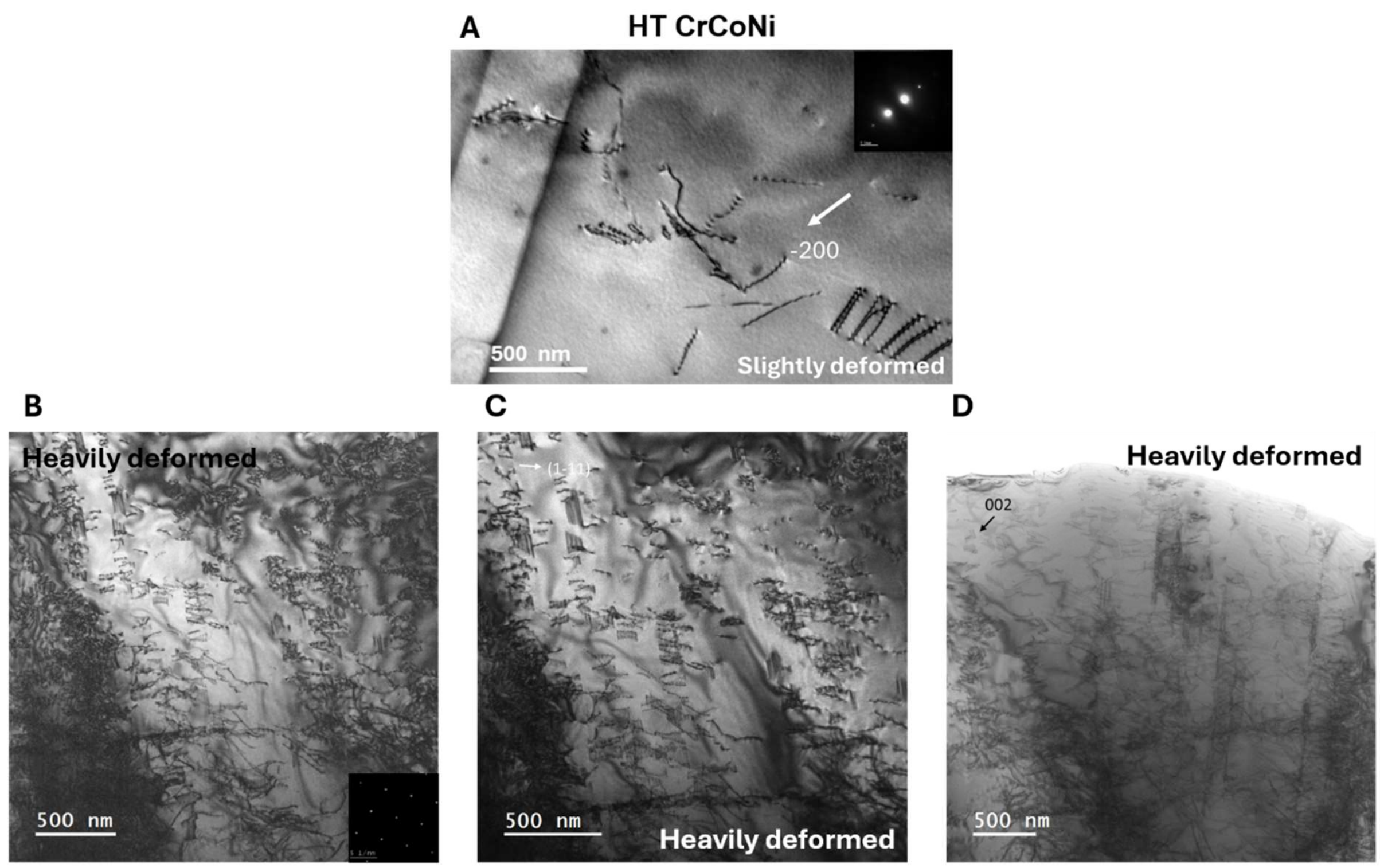

**Fig. S7 TEM analysis of HT CrCoNi samples.** A) Slightly deformed region of the tensile bar. B-D) Images from the heavily deformed regions.